\documentclass[preprint]{aastex631}

\usepackage{longtable,bm}
\usepackage{CJKutf8, color}
\usepackage{soul,multirow}

\renewcommand{\figurename}{Fig.}

\newcommand{\chandra}{\textit{Chandra}}
\newcommand{\lynx}{\textit{Lynx}}

\shorttitle{NS 1987A with 50+ years old}
\shortauthors{Dohi et al.}

\begin{document}

\title{Investigating Time Evolution of Thermal Emission from the Putative Neutron Star in SN 1987A for 50+ Years}

\correspondingauthor{Akira Dohi}
\email{akira.dohi@riken.jp}

\author[0000-0001-8726-5762]{Akira Dohi}
\affiliation{Department of Physics, Graduate School of Advanced Science and Engineering, Hiroshima University, Higashi-Hiroshima, Hiroshima 739-8526, Japan}
\affiliation{RIKEN Interdisciplinary Theoretical and Mathematical Sciences Program (iTHEMS), 2-1 Hirosawa, Wako, Saitama 351-0198, Japan}

\author{Emanuele Greco}
\affiliation{Anton Pannekoek Institute for Astronomy, University of Amsterdam, Science Park 904, 1098 XH Amsterdam, The Netherlands}
\affiliation{GRAPPA, University of Amsterdam, Science Park 904, 1098 XH Amsterdam, The Netherlands}
\affiliation{INAF-Osservatorio Astronomico di Palermo, Piazza del Parlamento 1, 90134 Palermo, Italy}
\affiliation{Universit\`a degli Studi di Palermo, Dipartimento di Fisica e Chimica, Piazza del Parlamento 1, 90134 Palermo, Italy}

\author{Shigehiro Nagataki}
\affiliation{Astrophysical Big Bang Laboratory (ABBL), RIKEN Cluster for Pioneering Research, 2-1 Hirosawa, Wako, Saitama 351-0198, Japan}
\affiliation{RIKEN Interdisciplinary Theoretical and Mathematical Sciences Program (iTHEMS), 2-1 Hirosawa, Wako, Saitama 351-0198, Japan}
\affiliation{Astrophysical Big Bang Group (ABBG), Okinawa Institute of Science and Technology Graduate University, 1919-1 Tancha, Onna-son, Kunigami-gun, Okinawa 904-0495, Japan}

\author{Masaomi Ono}
\affiliation{Institute of Astronomy and Astrophysics, Academia Sinica, Taipei 10617, Taiwan}
\affiliation{Astrophysical Big Bang Laboratory (ABBL), RIKEN Cluster for Pioneering Research, 2-1 Hirosawa, Wako, Saitama 351-0198, Japan}
\affiliation{RIKEN Interdisciplinary Theoretical and Mathematical Sciences Program (iTHEMS), 2-1 Hirosawa, Wako, Saitama 351-0198, Japan}

\author{Marco Miceli}
\affiliation{Universit\`a degli Studi di Palermo, Dipartimento di Fisica e Chimica, Piazza del Parlamento 1, 90134 Palermo, Italy}
\affiliation{INAF-Osservatorio Astronomico di Palermo, Piazza del Parlamento 1, 90134 Palermo, Italy}

\author{Salvatore Orlando}
\affiliation{INAF-Osservatorio Astronomico di Palermo, Piazza del Parlamento 1, 90134 Palermo, Italy}

\author{Barbara Olmi}
\affiliation{INAF-Osservatorio Astronomico di Palermo, Piazza del Parlamento 1, 90134 Palermo, Italy}

\begin{abstract} 

Observations collected with the Atacama Large Millimeter/submillimeter Array (ALMA) and analysis of broadband X-ray spectra have recently suggested the presence of a central compact object (CCO) in SN 1987A. However, no direct evidence of the CCO has been found yet. Here we analyze \chandra\ X-ray observations of SN 1987A collected in 2007 and 2018, and synthesize the 2027 \chandra\ and 2037 \lynx\ spectra of the faint inner region of SN 1987A. We estimate the temporal evolution of the upper limits of the intrinsic luminosity of the putative CCO in three epochs (2018, 2027 and 2037). We find that these upper limits are higher for higher neutron star (NS) kick velocities due to the increased absorption from the surrounding cold ejecta. We compare NS cooling models with both the intrinsic luminosity limits obtained from the X-ray spectra, and the ALMA constraints with the assumption that the observed blob of SN 1987A is primarily heated by thermal emission. We find that the synthetic \lynx\ spectra are crucial to constrain physical properties of the CCO, which will be confirmed by future observations in the 2040s. We draw our conclusions based on two scenarios, namely the non-detection and detection of NS by \lynx. If the NS is not detected, its kick velocity should be $\sim 700~{\rm km~s^{-1}}$. Furthermore, the non-detection of the NS would suggest rapid cooling processes around the age of 40 years, implying strong crust superfluidity. Conversely, in the case of NS detection, the mass of the NS envelope must be high.
\end{abstract}


\section{Introduction} \label{sec:intro}

 The explosion of the core-collapse supernova (SN) 1987A, which was confirmed through the detection of neutrinos on 23rd February in 1987~\citep{1987PhRvL..58.1490H,1987PhRvL..58.1494B}, is of great importance for understanding the physics of young central compact objects (CCOs). However, there has been no direct detection of the CCO of SN 1987A yet. With recent observational progresses, the first hints of the CCO emission have been reported by the Atacama Large Millimeter/submillimeter Array (ALMA) observations~\citep{2019ApJ...886...51C}. The authors suggested the existence of a warm dust blob hiding the CCO with (40--90)$~L_{\odot}$. For the explanation of the observed luminosity, several mechanisms have been argued, such as heating by the radioactive decay of $^{44}$Ti , magnetospherically-powered emission from the spin-down of a young pulsar, accretion-powered heating, and thermal (black-body) emission from the CCO~(see also Table 1 in \citealt{2020ApJ...898..125P}). Thus,  further investigation is needed in order to clarify the origin of the observed excess in the luminosity and therefore, to potentially identify the CCO emission.

 \cite{2020ApJ...898..125P} investigated the thermal emission scenario with their NS cooling models, concluding that thermal emissions from the CCO could reproduce the observed excess luminosity. Hence, multi-wavelength observations of SN 1987A can probe the NS model parameters, such as envelope properties, equation of state (EOS), the NS mass, and nucleon superfluid/superconductive models. They suggested that consistent cooling models of the possible NS in SN 1987A (NS 1987A hereafter) must have many light elements on the surface; the corresponding envelope mass is $M_{\rm env}\gtrsim10^{-9}~M_{\odot}$\footnote{Assuming that the degenerated electrons are dominant for supporting the outer crust and all nuclei are symmetric, one can get a relation~(e.g., Ref.~\cite{2021PhR...919....1B}):
 \begin{eqnarray}
M_{\rm env} = 1.9\times10^{-9}\rho_8g_{\rm s,14}^{-2}M_{\rm NS}~,
 \end{eqnarray}
 where $M_{\rm NS}$ is NS mass in a unit of $M_{\odot}$, $g_{s,14}$ is the surface gravity normalized by $10^{14}~{\rm cm~s^{-2}}$, and $\rho_8$ is the maximum density reached by the light elements in the crust normalized by $10^8~{\rm g~cm^{-3}}$. Thermal emission scenario requires $\rho_8\gtrsim1$ for the ALMA observation of SN 1987A~\citep{2020ApJ...898..125P}, implying $M_{\rm env}\gtrsim1.2\times10^{-9}~M_{\odot}$ for canonical NSs with $M_{\rm NS}=1.4~M_{\odot}$ and the radius of 12 km.} They also mentioned that the superfluidity in the ${}^1S_0$ state of neutrons in the crust may be weak in a certain range of $M_{\rm env}$ (see their Figure C1). Thus, the information of SN 1987A helps to constrain the NS models, which have still been uncertain despite several recent experiments and observations (e.g., \citealt{2016PhR...621..127L,2016ARA&A..54..401O}; for more recent constraints, see \citealt{2022PTEP.2022d1D01S} and reference therein.).

Thermal luminosities from a NS can be calculated from NS cooling theories, which describe how the NS cools down after its birth mainly by neutrino losses (at ages $t\lesssim10^5~{\rm yrs}$), because their mean free path is larger than the NS radius ~\citep{1983bhwd.book.....S}. Neutrinos are produced by many kinds of particle reactions inside the NS, e.g., modified/direct Urca processes, bremsstrahlung, and pair breaking and formation (PBF) in nucleon superfluid states. Hence, the time evolution of the thermal luminosity, which we call the cooling curve, is affected by the interior NS properties. 
Moreover, the envelope also affects the thermal luminosity through the thermal/electron conductivities and possible accretion and nuclear heating (for a review, see \citealt{2021PhR...919....1B}). Many works have investigated complex behaviors of cooling curves to explain many temperature observations of NSs~(for reviews, see, e.g., \citealt{2004ARA&A..42..169Y,2006NuPhA.777..497P}). In particular, the young NS of Cassiopeia A ($\sim340~{\rm yrs}$) is the first observational target to test the cooling theories for an early phase, and the modeling has been very successful indicating strong neutron superfluidity in the core ($T_{\rm cr, peak}\sim$  (5--6)$\times10^{8}~{\rm K}$, where $T_{\rm cr, peak}$ is the maximum superfluid transition temperature in all-density regions, ~\citealt{2011PhRvL.106h1101P,2011MNRAS.412L.108S}). While Cassiopeia A observations allow us to extract information on the superfluidity in the core, NS 1987A would provide relevant hints about the superfluidity in the crust, assuming thermal emission of the NS as the origin of the observed excess in ALMA observation. This is because of the difference of the thermal relaxation time ($t_w$) when the heat in the crust or core is completely transported to the surface, which is typically $t_w\gtrsim100~{\rm yrs}$ for the heat in the core while $t_w\lesssim100~{\rm yrs}$ in the crust~\citep{1994ApJ...425..802L,2001MNRAS.324..725G,2020A&A...642A..42S}. 
Therefore, SN 1987A is one of the few candidates to probe the crust superfluidity, which has still been unclear (but see also the recent study of ultra-cold atom gas by \citealt{2019NatSR...918477T}).

X-ray observations provide beneficial information of NS 1987A. For example, \cite{2021ApJ...908L..45G} analyzed \chandra\ and {\it NuSTAR} observations of SN 1987A and suggested that the spectra could be better explained by including a non-thermal component. The non-thermal component is most likely arising from a pulsar wind nebula (PWN) (but see also \citealt{2021ApJ...916...76A}). This scenario is further investigated by a recent follow-up paper \citep{gmo22}, which also provides additional information on the putative NS spin period and time derivative. Thus, although the direct X-ray observation of the NS 1987A has not been successful yet, our understanding on the properties of the NS continues to improve. 

Even in the case of a non-detection of the CCO, X-ray observations of SN 1987A can however provide upper limits for the thermal luminosity, which provide us with information of NS 1987A. 
Actually, \cite{2008AstL...34..675S} investigated consistent cooling models of NS 1987A with \chandra\, {\it XMM--NEWTON}, and {\it INTEGRAL} observations around 2000, which show $L^{\infty}_{\gamma,35}\lesssim0.2$, where $L^{\infty}_{\gamma,35}$ is the redshifted photon luminosity in units of $10^{35}~{\rm erg~s^{-1}}$. They concluded that such an upper limit for NS 1987A requires distinctive physical factors to cool around $t\simeq13~{\rm yrs}$, such as a strong crust superfluidity. In this work, we similarly constrain on the cooling curves by considering both ALMA and recent X-ray observations. Regarding the ALMA constraints, we assume that the CCO is an NS emitting thermal emission with $L^{\infty}_{\gamma,35}\sim1$, which is responsible for the heating of the observed blob, i.e., thermal emission scenario (see Section~\ref{sec:main} for details).
We also investigate future upper limits of the luminosity on the X-ray thermal emission, based on synthetic \chandra\ and \lynx\ spectra, exploiting the diagnostic power provided by the state-of-the-art MHD simulation of SN 1987A by \cite{oon20}.

This paper is organized as follows. In Section~\ref{sec:obs}, we describe the details of the analysis of recent \chandra\ observations, and we present the upper limits on the intrinsic luminosity of the NS. In Section~\ref{sec:setup}, we summarize the information of our NS cooling models, which include neutrino and photon cooling processes, superfluid models, and envelope models. In Section~\ref{sec:main},
we present our cooling models and compare them with the ALMA, \chandra\, and \lynx\ (future) observations. Then, we discuss the two cases of non-detection and detection of the CCO at the 2040s by a predicted future \lynx\ X-ray observation. We finally give concluding remarks in Section~\ref{sec:conc}.

\section{Analysis of the \chandra\ observations of SN 1987A}
\label{sec:obs}

Considering the young age of the system and the expected NS kick velocity of roughly $300-700$~km~s$^{-1}$ (see \citealt{oon20, 2019ApJ...886...51C}), we expect that the CCO lies in the internal area of SN 1987A, within a radius of $\sim 0.5''$ and the \chandra/ACIS Charge-Coupled Device (CCD) is the only X-ray detector currently able to spatially resolve it. In this paper, we consider 27 different \chandra\ observations performed in 2007 and 2018. We summarize the main information of all the observations in Table \ref{tab:obs}. 

\begin{table}[!ht]
    \centering
    \caption{Main characteristics of the Chandra observations}
    \begin{tabular}{c|c|c|c}
    Date & ObsID & PI& Exposure time (ks) \\
    \hline
    2007/03/11& 8523& Canizares & 30\\
    2007/03/12& 8537& Canizares & 13\\
    2007/03/13& 7588& Canizares & 27\\
    2007/03/18& 8538& Canizares & 21\\
    2007/03/19& 7589& Canizares & 25\\
    2007/03/20& 8539& Canizares & 25\\
    2007/03/21& 8542& Canizares & 18\\
    2007/03/24& 8487& Canizares & 29\\
    2007/03/27& 8543& Canizares & 31\\
    2007/03/28& 8544& Canizares & 19\\
    2007/03/29& 8488& Canizares & 32\\
    2007/03/31& 8545& Canizares & 20\\
    2007/04/01& 8546& Canizares & 31\\
    2007/04/17& 7590& Canizares & 35\\
    \multicolumn{3}{c}{Tot 2007} & 338 \\ 
\hline 
    2018/03/14& 20927& Park & 17 \\
    2018/03/15& 21037& Park & 30 \\
    2018/03/18& 21038& Park & 34 \\
    2018/03/19& 20322& Park & 15 \\ 
    2018/03/23& 21042& Park & 41 \\
    2018/03/25& 21043& Park & 29 \\
    2018/03/26& 21044& Park & 15 \\
    2018/03/27& 20323& Park & 27 \\
    2018/03/28& 21049& Park & 31 \\
    2018/03/29& 21050& Park & 18 \\
    2018/03/30& 21051& Park & 15 \\
    2018/03/31& 21052& Park & 30 \\
    2018/04/02& 21053& Park & 12 \\
    \multicolumn{3}{c}{Tot 2018} & 314\\
    \end{tabular}
    
    \label{tab:obs}
\end{table}

\subsection{Data reduction}

\begin{figure}[!ht]
    \centering
    \includegraphics[width=\textwidth]{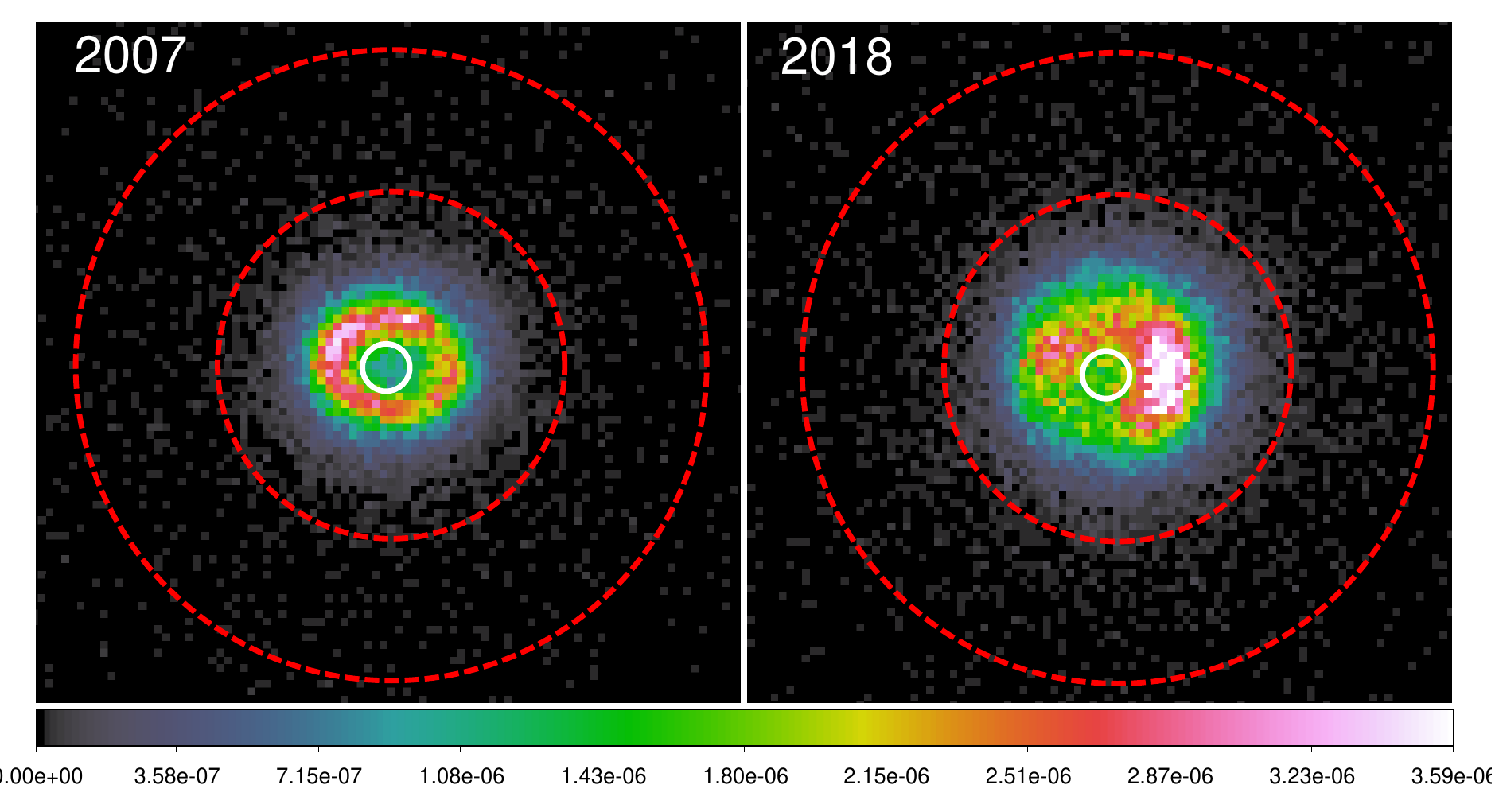}
    \includegraphics[width=\textwidth]{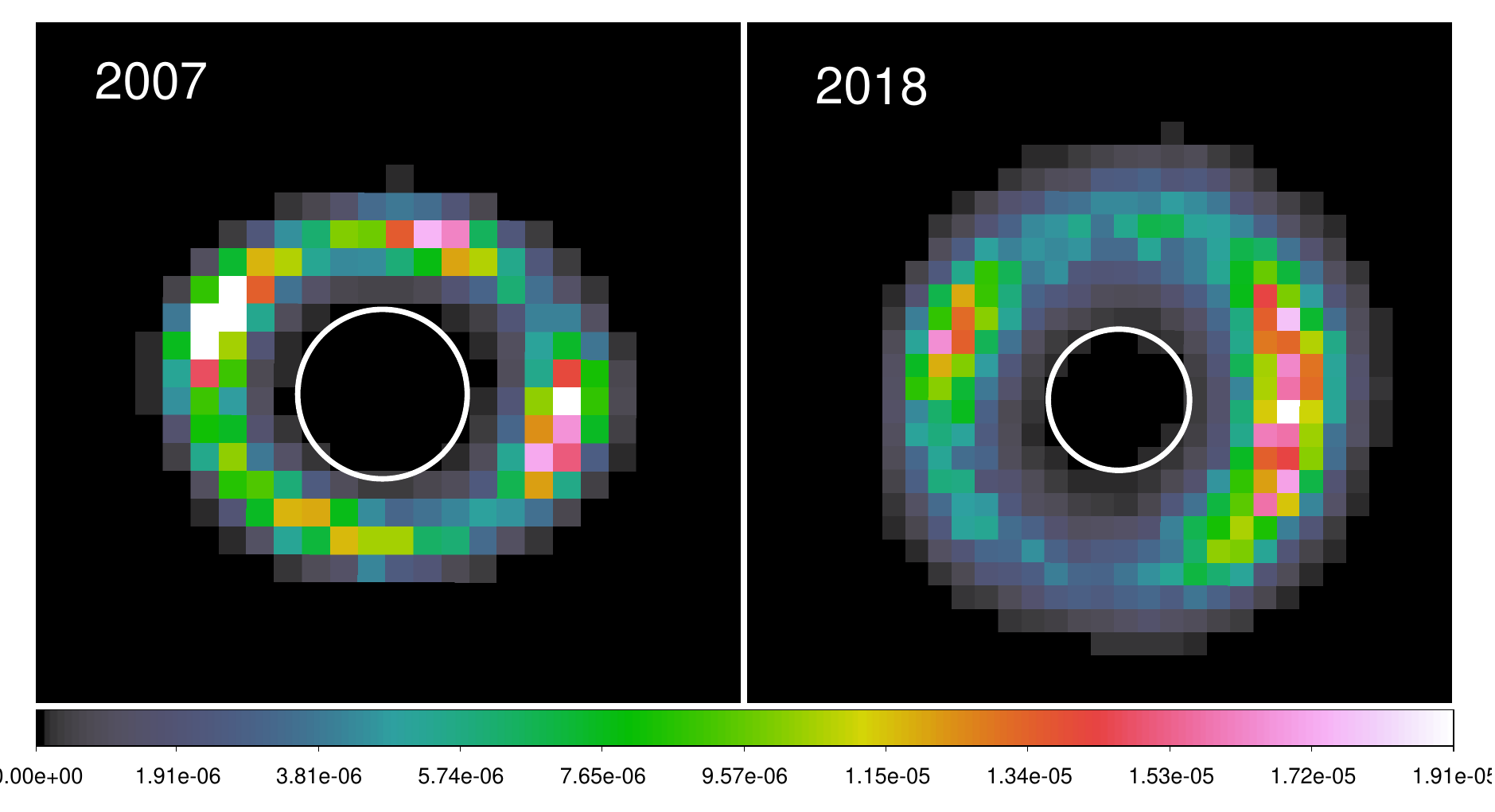}
    \caption{Broad (0.5-7 keV) \chandra/ACIS-S exposure-corrected count-rate maps of SN 1987A in 2007 and 2018, using a subpixel sampling (binsize of $0.1^\prime$). {\it Upper panels.} Source and background regions are shown in white and dashed red, respectively. {\it Lower panels.} Same as upper panels but the count-rate maps are deconvolved by the \chandra\ PSF through the Lucy algorithm.}
    \label{fig:chandra_imgs}
\end{figure}

We reduce the data following the standard procedure within the \emph{CIAO} software, version 4.12 with CALDB 4.9.2.1, through the task \texttt{chandra\_repro} (\citealt{ciao}). We merge observations performed in the same year (2007 and 2018) with \texttt{merge\_obs} to produce count-rate image of SN 1987A in the broad $0.5-7$ keV band (upper panels of Fig. \ref{fig:chandra_imgs}). We deconvolve the resulting images by the \chandra/ACIS-S point-spread-function (PSF) with the Lucy algorithm \citep{luc74,ric72} through the \texttt{arestore} task (lower panels of Fig. \ref{fig:chandra_imgs}). We considered a convergence criterion based on the self-similarity of the images after a given number of iterations, equal to 50 in our case.

The putative NS is expected to have a kick velocity \citep{oon20} which, projected on the plane of the sky, typically is $\lesssim 2000$ ${\rm km~s^{-1}}$ \citep{hll05,2018ApJ...856...18K}. Therefore, we select a $0.3''$  radius circle from the deconvolved count-rate maps and consider this as the region in which to look for possible radiation from the NS, as done by \citet{erl18}. The resulting source and background regions used for spectral extraction are shown in Fig. \ref{fig:chandra_imgs} in white and red, respectively. 

We extract the \chandra\ spectra from each observation with the \texttt{specextract} tool within \emph{CIAO}. The resulting spectra show $< 100$ counts. To improve spectral statistics, we combine all the spectra and the corresponding response and background files collected in the same epoch through the HEAsoft task \texttt{addascaspec} achieving reasonable error bars and binning. The corresponding combined response file is obtained by properly averaging between all the response files. This choice is justified by the small time range in which coeval observations are performed, $\sim 15$ days. Thanks to this approach, we obtain a unique spectrum for each epoch, with $\gtrsim 1000$ counts. The combined 2007 and 2018 spectra are then rebinned to 25 counts per bin. 

\subsection{Spectral analysis}

We fit the combined spectra  by adopting a model including a foreground absorption component (\texttt{TBabs} model in XSPEC; \citealt{arn96}), and an optically thin isothermal component emitting in non-equilibrium of ionization (\texttt{vnei} model in XSPEC). In the following, we refer to this \texttt{TBabs*vnei} model as 1-nei model. The total column density N$_{\rm{H}}$\footnote{Strictly speaking, this is an ``equivalent average” column density assuming Milky Way interstellar medium abundances, to have two different absorption components with Galactic and LMC abundances, respectively.} is fixed to $2.35 \times 10^{21}$ cm$^{-2}$ (\citealt{pzb06}). Temperature, emission measure, ionization parameter and the plasma metallicity are left free to vary.

\begin{figure*}
    \centering
    \includegraphics[width=.6\textwidth,angle=270]{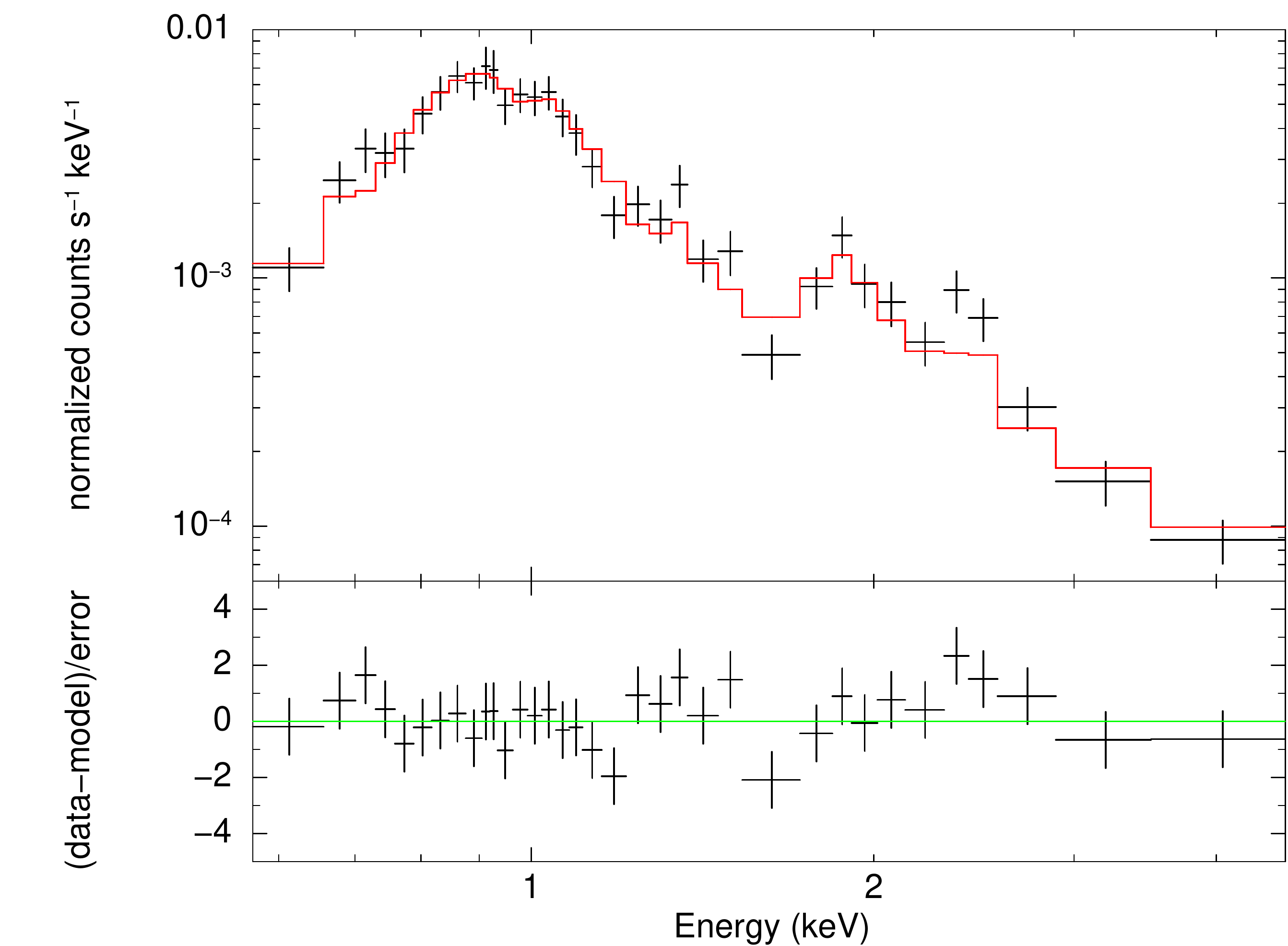}
    \includegraphics[width=.6\textwidth,angle=270]{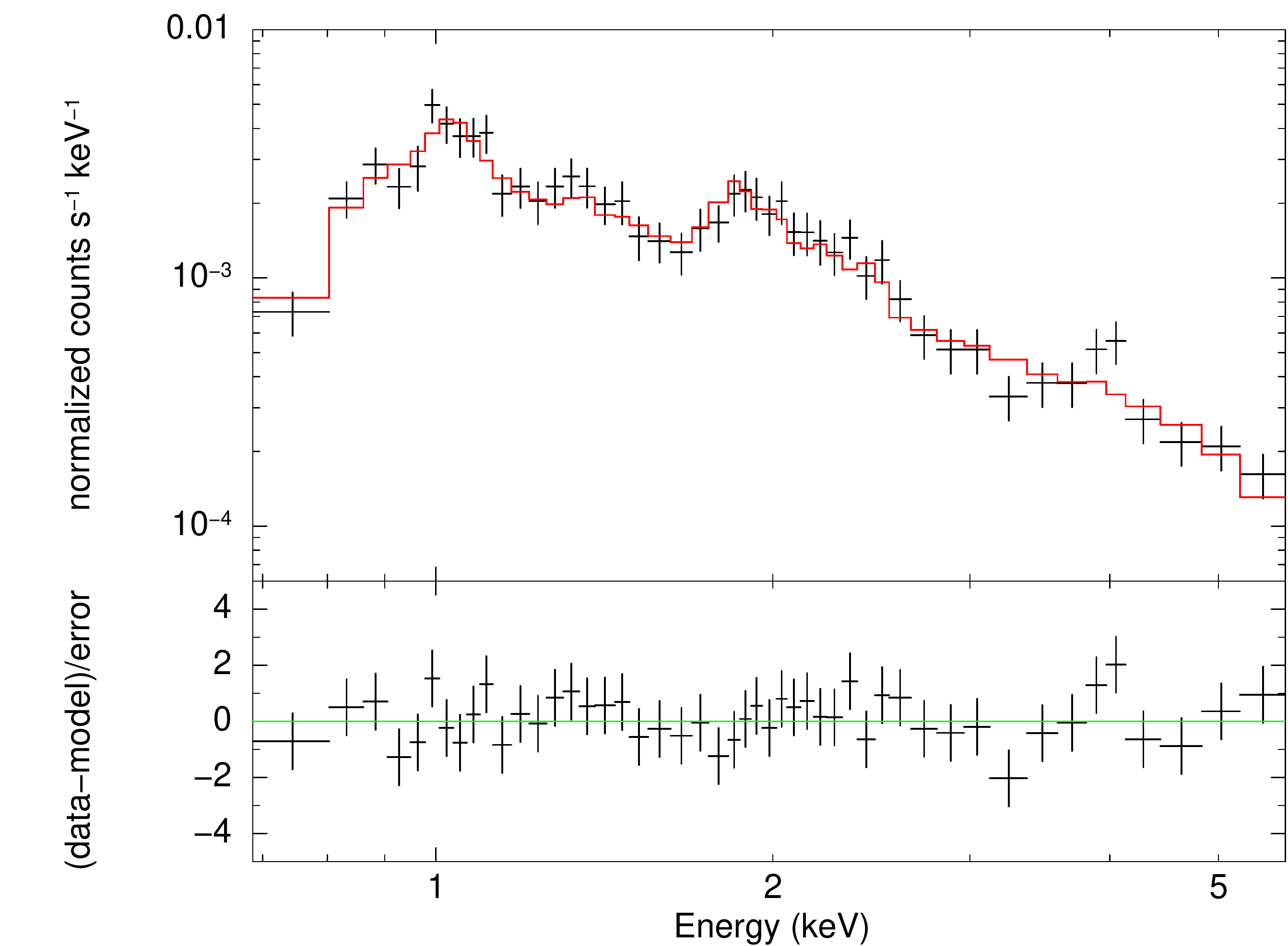}
    \caption{Combined \chandra/ACIS-S spectra (in black) extracted from the inner region of SN 1987A (shown in white in Fig \ref{fig:chandra_imgs}), with the corresponding best-fit model (in red) and residuals (bottom window of each panel). Data from 2007 and 2018 are shown, respectively, in the upper and bottom panels.}
    \label{fig:spec_chandra}
\end{figure*}

By adopting the 1-nei model, we obtain a $\chi^2 = 47.98$ with 31 degrees of freedom (d.o.f) in 2007 and a $\chi^2$ = 49.98 with 44 d.o.f. in 2018. Subsequently, we add another \texttt{vnei} component, which we refer to this \texttt{TBabs*(vnei+vnei)} as 2-nei model. Then, we obtain an improvement in the fit quality, achieving $\Delta \chi^2=-15$ in 2007 and $\Delta\chi^2=-17$ in 2018 with three additional parameters. In this 2-nei model scenario, we force the metallicities of the two components to be equal. We note that in both the models the ionization age $\tau$ is poorly constrained. This is an effect of the low statistics of the spectra. In any case, as discussed in the next subsection, our results do not depend on this particular value. The 2007 and 2018 \chandra\ spectra with the corresponding best-fit model and residuals are shown in Fig. \ref{fig:spec_chandra}. Details of the best-fit parameters are shown in Table \ref{tab:fit_observed}.

\begin{table}[!ht]
    \centering
    \caption{Best-fit parameters of the 2007 and 2018 spectra}
    \begin{tabular}{c|c|c|c|c}
    & \multicolumn{2}{c|}{2007}& \multicolumn{2}{c}{2018} \\ 
    \hline 
    Parameter & 1-nei model & 2-nei model & 1-nei model & 2-nei model \\
    \hline
    N$_{\rm{H}}$ (10$^{21}$ cm$^{-2}$) & \multicolumn{4}{c}{2.35 (fixed)} \\
    \hline
    kT$_{\rm{1}}$ (keV) & 0.81$_{-0.16}^{+0.26}$& 0.81$_{-0.16}^{+0.26}$ & 1.9$^{+0.3}_{-0.2}$  & 0.7$_{-0.3}^{+0.2}$ \\
    Abundance$_1$ & 0.30$_{-0.09}^{+0.13}$& 0.9$_{-0.3}^{+1.6}$& 0.19$_{-0.05}^{+0.07}$& 0.24$_{-0.10}^{+0.16}$ \\
    $\tau_{\rm{1}}$ (10$^{11}$ s cm$^{-3}$) & 3.8$_{-0.8}^{+1.3}$&0.9$\pm 0.4$ & 0.41$_{-0.10}^{+0.15}$& $> 1$ \\
    Emission measure$_1$ (10$^{58}$ cm$^{-3}$) & 0.52$_{-0.15}^{+0.21}$& 0.23$_{-0.14}^{+0.37}$ & 1.1$\pm 0.2$ & 1.0$\pm 0.6$ \\
    \hline
    kT$_{\rm{2}}$ (keV) & / & 1.9$_{-0.5}^{+2.2}$ &/ &2.6$_{-0.5}^{+0.9}$ \\
    Abundance$_2$ &/ & =Abundance$_1$ & / & =Abundance$_1$ \\
    $\tau_{\rm{2}}$ (10$^{11}$ s cm$^{-3}$) & / & $> 1 $&/ & 0.8$_{-0.3}^{+0.5}$\\
    Emission measure$_2$ (10$^{58}$ cm$^{-3}$) & / &  0.23$_{-0.11}^{+0.3}$ &/ & 0.6$_{-0.2}^{+0.3}$\\
    \hline
    $\chi^2$ (d.o.f.) & 47.98 (31) & 32.78 (28) & 49.98 (44) & 32.99 (41)\\
    \end{tabular}
    
    \textbf{Notes.} Error bars are at 90$\%$ confidence level.
    \label{tab:fit_observed}
\end{table}

\subsection{Absorption from cold ejecta}
\label{sec:absorption}
To estimate the absorption pattern due to the cold ejecta, we exploit the magnetohydrodynamic (MHD) simulation of SN 1987A from \cite{oon20}. The complete details on the procedure we adopted to model the X-ray absorption form the cold ejecta are given in Appendix B of \cite{2021ApJ...908L..45G} and in Sect. 2.1 of \cite{gmo22}, thus here we only recall the main steps. Values of temperature, density, chemical composition and ionization age of the plasma are associated with each cell of the 3D spatial domain of the MHD model. We extract ionic density, chemical abundances and electronic temperature from each cell of the 3D domain and include them in the spectral analysis through the \texttt{vphabs} component available in XSPEC, which provides the absorbing power due to the photoelectric effect. We repeat the procedure for every year considered in this work (2007, 2018, 2027, 2037) also taking into account the proper motion of the putative CCO, whose kick velocity is provided by the MHD model under the assumption of momentum conservation (see \citealt{2020ApJ...888..111O} for details). We consider the module of the kick velocity, $v_{\rm kick} = 300$ ${\rm km~s^{-1}}$, provided by the model as a lower limit (\citealt{oon20}).
Moreover, from the analysis of ALMA observation, \cite{2019ApJ...886...51C} measured an offset between a dust blob (possibly related to the NS) and the location of the progenitor star estimated by \cite{alf18}, and derived an upper limit to the kick velocity of $700~{\rm km~s^{-1}}$. To maintain our analysis as general as possible, we estimate the ejecta absorbing power by assuming three possible kick velocities: 300 ${\rm km~s^{-1}}$, 500 ${\rm km~s^{-1}}$ and 700 ${\rm km~s^{-1}}$. 

We note that the three different kick velocities imply different linear momentum of the NS. Consequently, a different distribution of the ejecta for the three cases should be expected on the basis of the conservation of momentum. On the other hand, the linear momentum of the NS reflects the total linear momentum of the ejecta, i.e. the difference between the linear momentum of the two portions of the ejecta that propagate in the two opposite directions defined by the bipolar SN explosion assumed in the model of SN 1987A. This difference is much smaller (by a factor of ~ 20) than each of the momentum of these two portions of ejecta. An increase of a factor of two in the linear momentum of the NS (corresponding to increasing its kick velocity by a factor of 2) would imply very small changes in the total ejecta distribution. In light of this, we considered the NS kick velocity calculated from the model on the basis of exact momentum conservation ($\sim 300{\rm km~s^{-1}}$) and explored the possibility of higher kick velocities up to the value estimated by (\citealt{2019ApJ...886...51C}; namely 700 ${\rm km~s^{-1}}$) without the need to run new MHD simulations (which would require a very high computational cost).

\begin{figure}[!ht]
    \centering  
    \includegraphics{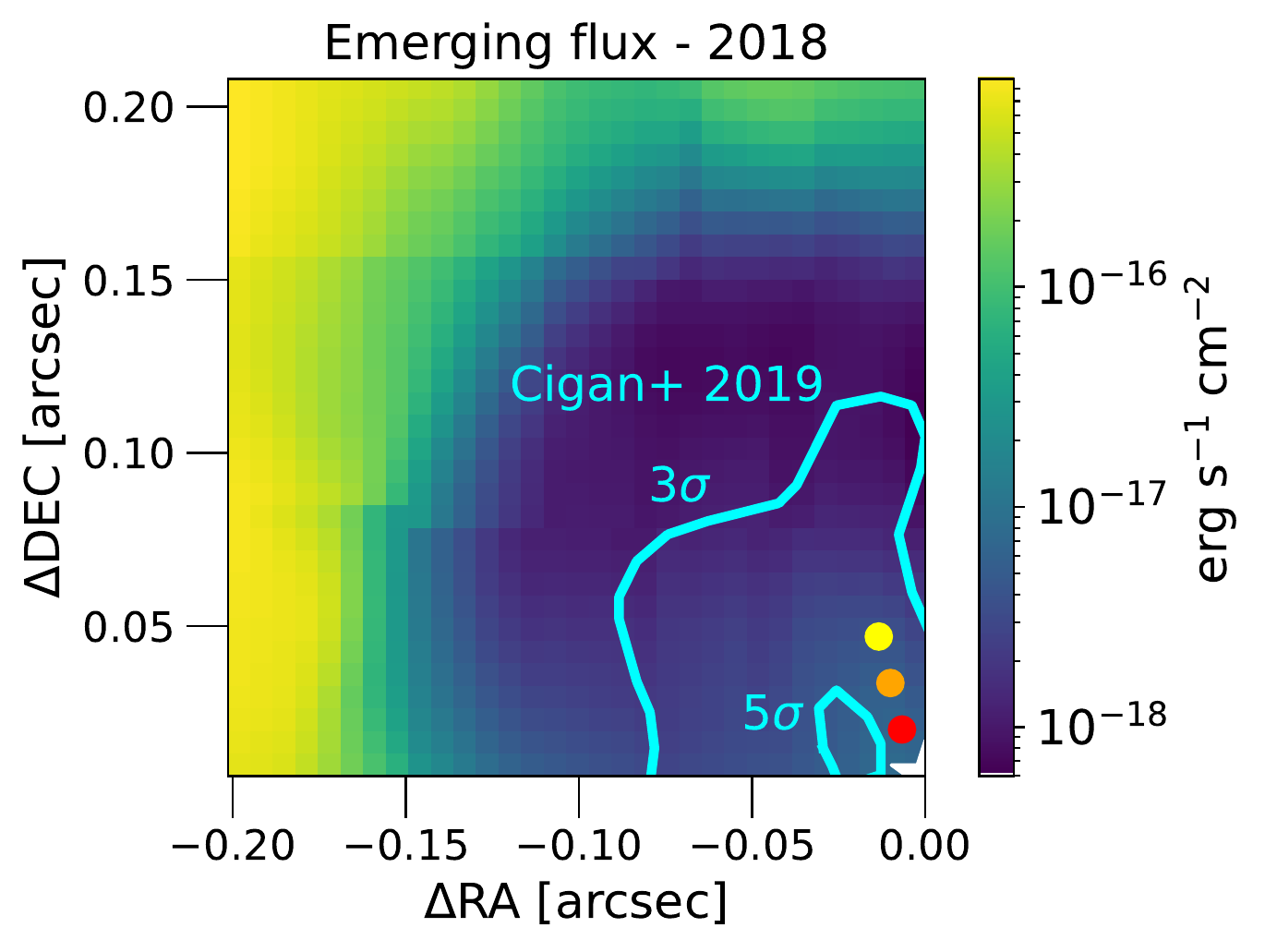}
    \caption{Map of the emerging flux for year 2018. The origin is set at the assumed center of explosion (bottom right corner). Red, orange and yellow dots indicate the position of the putative NS assuming the direction of the kick according to the MHD model of \cite{oon20} and a kick velocity of 300 ${\rm km~s^{-1}}$, 500 ${\rm km~s^{-1}}$ and 700 ${\rm km~s^{-1}}$, respectively. Cyan contours trace the 679 GHz dust emission at 3$\sigma$ and 5$\sigma$, as reported in Fig. 3 by \citet{2019ApJ...886...51C}.} 
    \label{fig:emerging_flux}
\end{figure}

We built a map (Fig.~\ref{fig:emerging_flux}) showing the bolometric emerging flux in the northeastern region of SN 1987A, where the NS is expected to be according to the MHD model. The remnant is oriented according to the inclination of the dense equatorial ring as found from the analysis of optical data (e.g., \citealt{2005ApJS..159...60S}). To each pixel of Fig. \ref{fig:emerging_flux} is associated a value of flux, expressed in erg~s$^{-1}$~cm$^{-2}$, estimated by convolving the redshifted emission irradiated by a 11.57 km radius and 1.4 M$_{\odot}$ NS with the ejecta absorption. Since the density and the chemical distribution change pixel by pixel, the corresponding emerging flux varies accordingly. 
In particular, darker areas mark lower emerging flux and, subsequently, higher ejecta absorbing power. Red, orange and yellow dots show the position of the putative NS assuming a kick velocity of 300 ${\rm km~s^{-1}}$, 500 ${\rm km~s^{-1}}$ and 700 ${\rm km~s^{-1}}$, respectively. 
The figure also shows cyan contours corresponding to the 679 GHz dust emission at $3\sigma$ and $5\sigma$ reported in Fig. 3 of \citet{2019ApJ...886...51C}, which are in good agreement with the findings in this paper. 
Fig. \ref{fig:emerging_flux} clearly shows that considering an higher kick velocity leads to a NS embedded in more effectively absorbing material, with the peak of the absorption corresponding to a shell at $\sim 0.1^{\prime}$ from the center of the explosion.

\subsection{Upper limits of the bolometric redshifted luminosity}

We include in our model a \texttt{bbodyrad} component to look for thermal (blackbody) radiation stemming from the putative NS. Two effects influence the black-body spectral shape: i) the gravitational redshift z; ii) the absorption from the interstellar medium along the line of sight and from the inner ejecta lying in the very hearth of SN 1987A. Besides, the model of NS atmosphere also changes the spectral shape through radiative thermal conductivity, which is explained in Section \ref{sec:setup}. The gravitational redshift can be estimated given the radius and the mass of the NS. The foreground absorption from the interstellar medium is already included in our model with the $\texttt{TBabs}$ component as fixed to be $N_H=2.35\times10^{21}~{\rm cm}^{2}$~\citep{pzb06}. The absorption due to the inner ejecta requires a bit more effort and it is extensively discussed in Sect. \ref{sec:absorption}.

We add the absorbed and redshifted black-body spectrum to our 2-nei model in both years. In the following, we refer to this model as \emph{bb model} \footnote{In XSPEC terms, this model reads as \texttt{TBabs*(vnei+vnei+vphabs(zashift(bbodyrad)))}. Note that the order of the \texttt{vphabs} and \texttt{zashift} matters, since the blackbody radiation is first redshifted and then absorbed.}. Since the characteristic dimension of the inner region of SN 1987A is smaller than the \chandra\ PSF, we correct the normalization of the \texttt{bbodyrad} component to take into account the fractional encircled energy (see Fig. 4.6 at the \href{https://cxc.harvard.edu/proposer/POG/pdf/MPOG.pdf}{Chandra OG website}). In our case, the peak of the absorbed blackbody is at roughly 4 keV, which, for a region with a radius of $0.3^{\prime}$, requires a correction factor $ \rm{f}_{\rm{PSF}} \sim$ 0.5. Therefore, we correct the normalization of the \texttt{bbodyrad} component by $ \rm{f}_{\rm{PSF}}$ before translating its best-fit value into the radius of the NS.

All the parameters relative to the redshift and ejecta absorption components are kept frozen during the fitting procedure. We find no significant improvement in the description of the data with this \emph{bb model}, even by considering different kick velocities and/or masses of the NS, indicating that there is no evidence for any thermal emission stemming from the NS. Given the non-detection of the NS radiation, we can only put upper limits on its flux. The most constraining upper limits come from the 2018 data, since the absorbing power of the cold ejecta decreases with time due to their expansion (see \citealt{gmo22} for a public repository of the absorption components from 2001 to 2037). Therefore, in the following, we focus on these more recent data sets.

As mentioned above, fitting the data with a 2-nei model or the \emph{bb model} leads to identical values of $\chi^2$. To investigate the upper limit on the luminosity of the putative NS we estimate the maximum blackbody luminosity compatible with the observed spectrum at 90\% (1.6$\sigma$) and 99.7\% (3$\sigma$) confidence levels. We adopt the following approach: we fix all the parameters corresponding to the emission of the NS, i.e. the redshift, the normalization and the temperature; we fix the mass and the radius of the emitting blackbody; we keep free to vary all the parameters of the 2-nei components, as shown in Table \ref{tab:fit_observed}; we fit the data by increasing the temperature of the blackbody until a $\Delta\chi^2 = 2.706$ or $\Delta\chi^2 = 9$ is achieved, evaluated with respect to the best-fit 2-nei model.  With this approach, we obtain the maximum temperature allowed for the given configuration of the \emph{bb model}. Since the luminosity of a black-body, at fixed radius, depends only on the temperature we also infer the upper limits of the bolometric redshifted luminosity of the NS. We repeat this procedure for the three kick velocities (300 ${\rm km~s^{-1}}$, 500 ${\rm km~s^{-1}}$ and 700 ${\rm km~s^{-1}}$) and three masses (1.2 M$_{\odot}$, 1.4 M$_{\odot}$ and 1.6 M$_{\odot}$) of the NS. The resulting 2018 upper limits are shown in Tables \ref{tab:upper_limits_1.2}, \ref{tab:upper_limits_1.4} and \ref{tab:upper_limits_1.6}. 

The 2018 upper limits we find are in good agreement with the ones reported by \citet{alf18}. However, as we show in the next sections, these values do not allow to exclude any of the possible cooling curves relative to the thermal emission of the putative NS. Therefore, we investigate how the luminosity upper limit would change in the upcoming years. Since the absorbing power of the ejecta decreases with time, we expect the luminosity upper limit to decrease as well.

We need to synthesize reliable future X-ray spectra in order to robustly estimate time variation of the upper limits. The MHD model by \citet{oon20} predicts that the flux arising from the SNR in the inner faint area of SN 1987A is constant between 2018 and 2037. Therefore, we synthesized the 2027 \chandra/ACIS-S and 2037 \lynx/X-ray Microcalorimeter (LXM) spectra of the inner area of SN 1987A, starting from the best-fit 2-nei model obtained for the 2018 \chandra/ACIS-S data (Table \ref{tab:fit_observed}). The luminosity upper limits also depend on the assumed exposure time, since the statistics affects the sensitivity of the spectra to the additional \texttt{bbodyrad} component. Therefore, we synthesize the 2027 and 2037 spectra assuming two different exposure times: 200 ks, similar to the \chandra/ACIS-S observations performed in 2018, and 1Ms, typical of large program observations. The resulting spectra are then analyzed by following the same procedure described for the 2018 data: we find the 2-nei best-fit model and then we estimate the 90\% and 99.7\% upper limits on luminosity through the \emph{bb model}. It is worth noticing that the \emph{bb model} used to fit the synthetic spectrum takes into account the variation of local absorption by the ejecta. The luminosity upper limits for 2027 and the 2040s are shown in Tables \ref{tab:upper_limits_1.2}, \ref{tab:upper_limits_1.4} and \ref{tab:upper_limits_1.6}.
It is noted that \lynx\ will not likely be launched in 2037, but our conclusion is qualitatively unchanged as long as \lynx\ is launched in the 2040s, as shown in subsection 4.3 for details. From such a standpoint, we dare to mostly label the launched years of \lynx\ as ``2040s", not 2037 since here.

\begin{figure*}[!h]
    \centering
    \includegraphics[width=\textwidth]{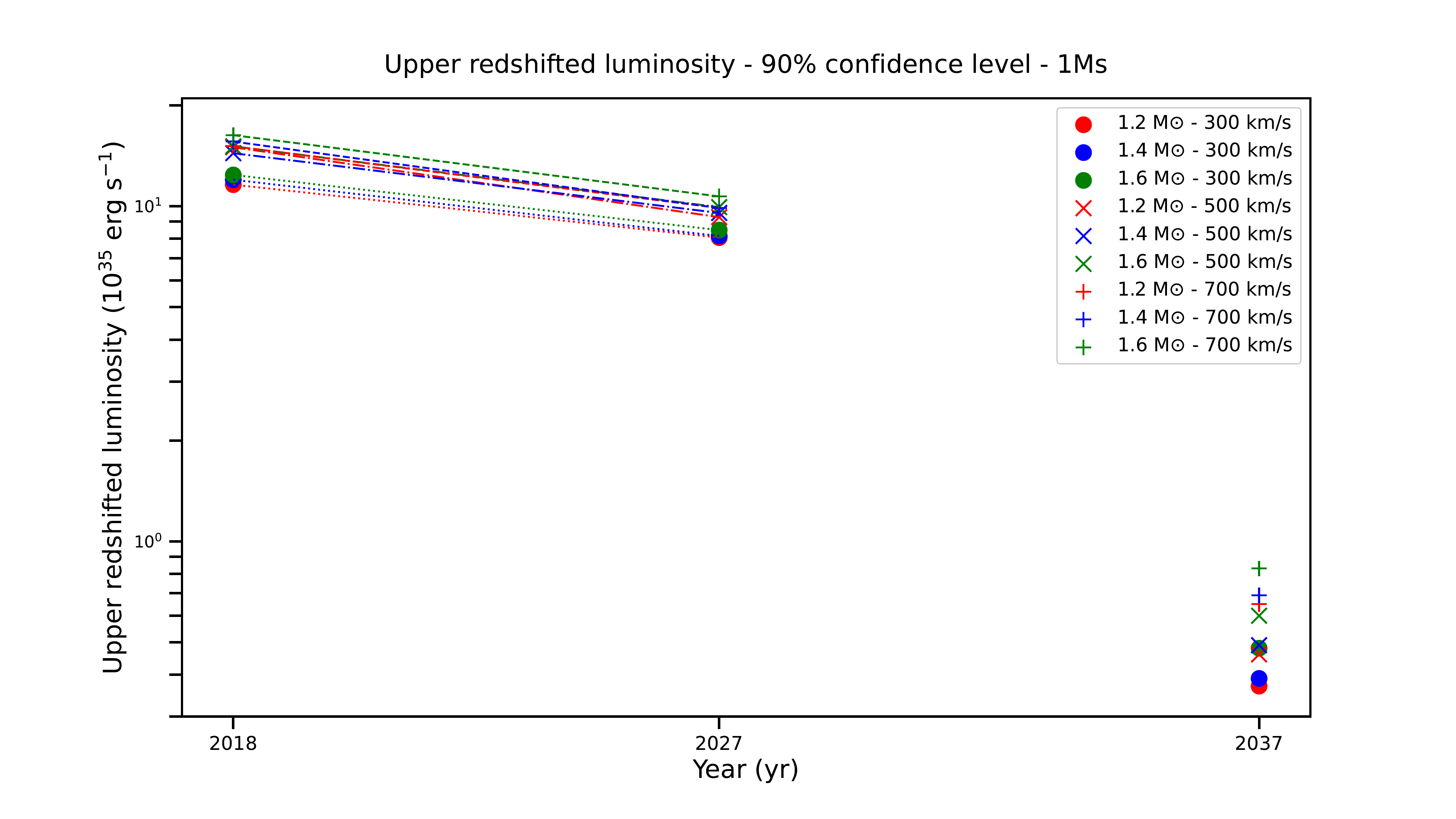}
    \caption{Evolution with time of upper limits of redshifted (bolometric) luminosity estimated at 90\% confidence level. The synthetic \chandra\ and \lynx\ spectra are produced assuming an exposure time of 1 Ms. Colors red, blue and green indicate a NS mass of 1.2 M$_{\odot}$, 1.4 M$_{\odot}$ and 1.6 M$_{\odot}$, respectively. The circles, crosses and pluses indicate a kick velocity of 300 ${\rm km~s^{-1}}$, 500 ${\rm km~s^{-1}}$ and 700 ${\rm km~s^{-1}}$, respectively. Upper limits derived from spectra collected with the same instrument (namely \chandra\,) are linked with lines.
    }
    \label{fig:upper_limits_90_1Ms}
\end{figure*}

We show the time evolution of the upper limits at 90\% confidence level, measured from the 1 Ms synthetic spectra, in Fig.~\ref{fig:upper_limits_90_1Ms} to better highlights the time evolution of these quantities. Other scenarios considering 99.7\% confidence level and 200ks exposure time provide less constraining upper limits and are shown in Appendix A (Figures \ref{fig:upper_limits_90_200ks}, \ref{fig:upper_limits_99_200ks}, \ref{fig:upper_limits_99_1Ms}).
The time evolution shown in Fig. \ref{fig:upper_limits_90_1Ms} clearly shows a decreasing trend. This is well expected, given that our reference MHD model predicts an expansion of the internal ejecta of SN 1987A, leading to less efficient photoelectric effect and, finally, to lower upper limits. The 1.2 M$_{\odot}$ - 300 ${\rm km~s^{-1}}$ is the scenario which shows the lowest upper limit, in each of the years considered. 
This is due both to the cocoon-like distribution of the ejecta, and to the gravitational redshift of the blackbody radiation: lower mass implies lower redshift. Since the X-ray flux due to shock-heated plasma is higher at energies $\lesssim 2$ keV, the more the blackbody radiation is redshifted, the higher is the flux hidden below the observed spectrum. Therefore, the redshift has a more important role than the size of the NS in the upper limit measure, which changes of just few percentages from the scenario with 1.6 M$_{\odot}$ (11.45 km radius) to the 1.2 M$_{\odot}$ one (11.65 km radius). 

We also notice a significant decrease (roughly a factor of 30) of the upper limits by \lynx\,, with respect to \chandra. We find the absorbed flux to increase by a factor of $\sim 3$ in that period, while the additional difference of a factor $\sim 10$ is due to the instrument used for the synthesis of the spectra. In fact, the high sensitivity of the \lynx\ X-ray microcalorimeter is beneficial to the estimate of the upper limits, since small variations of the \emph{bb model} spectrum lead to noticeable increase in the $\Delta\chi^2$ value. Subsequently, when \lynx\ will be operating we will be able to better discern between the various blackbody scenarios even if the CCO will not be detected yet.

Our study of the parameter space of the \emph{bb model} not only provides future upper limits for the bolometric luminosity, but it also may be used as a constrain on the NS or SNR characteristics in case of a direct detection of the compact object. For instance, if in the 2040s a successfull detection of the CCO will be performed, we would have constrain on its kick velocity, mass or radius.

\section{Cooling Models}
\label{sec:setup}

Thermal evolution of isolated NSs can be divided into two cooling stages: the first period is called the neutrino cooling era, in which neutrino processes working inside NSs are dominant in their thermal evolution. This era lasts until $t\sim10^5~{\rm yrs}$ after the formation of NSs, hence the putative NS 1987A lies in this era. Once the neutrino cooling era is over, a so-called photon cooling era begins, since the neutrino luminosity becomes lower than the photon luminosity due to the low internal temperature of the NS. To properly describe such a long-term thermal evolution with cooling models, we need a realistic EOS, cooling processes, superfluid/superconductive models, and envelope models. 

To describe the NS structure and composition, we adopt the widely-used Akmal-Pandharipande-Ravenhall (APR) EOS, which is built based on realistic two-body interaction and phenomenological three-body interaction~\citep{1998PhRvC..58.1804A}. The maximum mass is $2.183~M_{\odot}$, which agrees with Shapiro-delay based mass measurements of pulsars~\citep{2010Natur.467.1081D,2013Sci...340..448A,2020NatAs...4...72C}. 
If the proton fraction inside the NS exceeds 1/9 (without muons), the Direct Urca (DU) process works because of momentum conversation and cool the NS rapidly, whose threshold central density or mass naturally depends on the EOS \citep{1981PhLB..106..255B,1991PhRvL..66.2701L}. 
In the APR EOS, the threshold mass of the DU process is $1.96~M_{\odot}$.
Since current predictions from observations of  light-curves in  SN 1987A indicate $1.22\le M_{\rm NS}/M_{\odot}\le 1.62$~\citep{1988ApJ...330..218W,1988A&A...196..141S,2019A&A...624A.116U,2020ApJ...890...51E,2020ApJ...898..125P}, in this work we have limited the NS mass to the range $1.2$--$1.6\,M_{\odot}$, which then excludes the possibility to have DU processes.

The radii of 11.65, 11.57, and 11.45~km correspond to NSs with masses of $1.2,1.4$, and $1.6~M_{\odot}$ respectively, coherently with the X-ray analysis described in Section~\ref{sec:obs}.
We note that, under such a minimal cooling scenario~\citep{2004ApJS..155..623P}, the influence of the EOS uncertainties on cooling curves is not so large compared with other model parameters~(e.g., \citealt{2004ARA&A..42..169Y,2017IJMPE..2650015L}). 
 
In the neutrino cooling era, the neutrino emission processes that dominate in most NSs are so-called slow and medium cooling processes. The latter arises because of nucleon superfluidity. The former mainly includes the modified Urca (MU) process and bremsstrahlung in nucleon-nucleon collisions~\citep{1995A&A...297..717Y}. The emissivity of slow cooling processes is approximately expressed as follows:
\begin{eqnarray}
\epsilon^{\rm Slow}_{\nu} \approx 10^{19-21}\left(\frac{\rho_{\rm B}}{\rho_{\rm nuc}}\right)^{1/3} T_9^8 ~~{\rm erg~cm^{-3}~s^{-1}}\,,
\end{eqnarray}
where $\rho_{\rm B}$ is the baryon density, $\rho_{\rm nuc}\simeq2.8\times10^{14}~{\rm g\,cm^{-3}}$ is the nuclear saturation density, and $T_9$ is the temperature in units of $10^9~{\rm K}$. For the medium cooling processes, the pair breaking and formation (PBF) process, which is the latent heat released in making nucleons pair, works in accordance with the nucleon superfluid state. In terms of the type of superfluidity, neutrons in the inner crust and protons in the core become in the singlet state (${}^1S_0$) and neutrons in the core become in the triplet state (${}^3P_2$) if $T<T_{\rm cr}$, where $T_{\rm cr}$ is the superfluid transition temperature. The emissivity of PBF processes is approximately expressed as follows~\citep{1995A&A...297..717Y}:
\begin{eqnarray}
\epsilon^{\rm PBF}_{\nu} \approx 10^{21-22}\left(\frac{\rho_{\rm B}}{\rho_{\rm nuc}}\right)^{1/3} T_9^7\,\tilde{F}_i\!\left(\frac{T}{T_{\rm cr}}\right) ~~{\rm erg~cm^{-3}~s^{-1}}\,,
\end{eqnarray}
where $\tilde{F}_i(T/T_{\rm cr})$ is the control function
with $T/T_{\rm cr}$, which depends on the state of nucleons ($i=s$ for ${}^1S_0$ and $i=t$ for ${}^3P_2$, \citealt{1999A&A...343..650Y}):
\begin{eqnarray}
\tilde{F}_s = y^2\int_0^{\infty} dx\frac{z^4}{1+e^z},~~~~~~
\tilde{F}_t = \frac{1}{4\pi}\int d\Omega~y^2\int_0^{\infty}dx \frac{z^4}{1+e^z}~,
\end{eqnarray}
where $y=k_iT_{\rm cr}/T$, $z=\sqrt{x^2+y^2}$ and $\int d\Omega$ denotes the angle averaging procedure. $k_i$ is the conversion factor between $T_{\rm cr}$ and the gap $\Delta$ for each state of nucleons. $\tilde{F}_i$ reaches its maximum value (unity) for $T\sim 0.5 T_{\rm cr}$ and zero for $T\lesssim~0.2 T_{\rm cr}$~\citep{2004ApJS..155..623P}. Since the PBF process occurs through the vector or the axial channel, their coupling constants affect the emissivities. We mostly adopt values from \citet{2009ApJ...707.1131P}, where the axial channel is dominant for the PBF process~\citep{2008PhRvC..77f5808K}. Generally, the PBF emissivity is higher than that of slow cooling processes in young NSs. It is noted that the crust superfluidity basically decreases the neutron specific heat, which also affects the cooling curves of young NSs~\citep{2009ApJ...707.1131P}.

\begin{figure}[t]
    \centering  \includegraphics[width=0.7\linewidth]{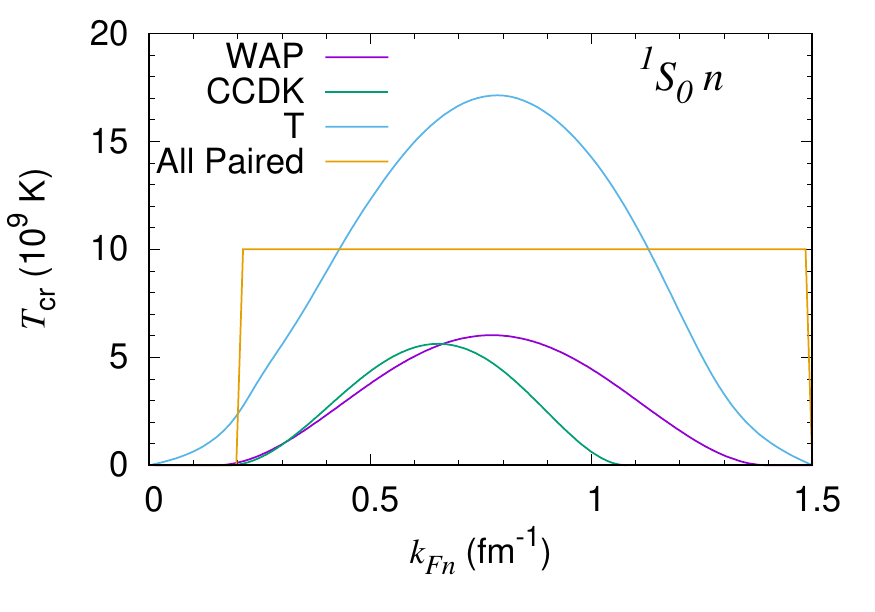}
    \caption{Adopted crust superfluid models, i.e., critical temperature as a function of Fermi wave number of neutrons.}
    \label{fig:sfn}
\end{figure}

Depending on the kind of superfluidity/superconductivity, the typical NS age when the superfluid effect significantly contributes to their cooling curves is varied. For the ${}^1S_0$ neutrons, the NS age where the superfluid effect can be clearly seen in cooling curves is 10--100 yrs. For the ${}^1S_0$ protons and ${}^3P_2$ neutrons, on the other hand, it is more than 100 yrs~(see \citealt{2015PhRvC..91a5806H}). As pointed out by \cite{2009ApJ...707.1131P,2020ApJ...898..125P}, the most uncertain factor of superfluidity for NS 1987A is the strength of ${}^1S_0$-neutrons pairing. To examine its model dependence of cooling curves, we adopt various superfluid models of ${}^1S_0$-neutrons as we show in Fig.~\ref{fig:sfn}; CCDK~\citep{1993NuPhA.555...59C}, WAP~\citep{1993NuPhA.555..128W}, T~\citep{1984PThPh..71.1432T}, and All Paired which assumes $T_{\rm cr}=10^{10}~{\rm K}$ in $k_{F_n}<1.48~{\rm fm}^{-1}$. For the  ${}^3P_2$-neutrons superfluidity, we adopt the model ``a" in \cite{2004ApJS..155..623P} multiplied by 0.59 ($T_{\rm cr,peak}=5.9\times10^{8}~{\rm K}$, where $T_{\rm cr,peak}$ is the maximum critical temperature), which matches the recent cooling observation of the CCO detected in Cassiopeia A (see also Figure 15 in \citealt{2019MNRAS.484..974W}). Compared with neutron superfluidity, the influence of proton conductivity on cooling curves under the minimal cooling scenario is much smaller~\citep{2015PhRvC..91a5806H}, so we fix the model as CCDK~\citep{1993NuPhA.555...59C}. 

In the photon cooling stage, the photon luminosity becomes higher than neutrino luminosity because the temperature dependence is larger for the latter. The photon luminosity of black-body emission can be expressed as follows:
\begin{eqnarray}
L_{\gamma}(r=r_s) = 7\times 10^{36}~{\rm erg~s^{-1}}\left(\frac{r_s}{10~{\rm km}}\right)^2T^4_{{\rm eff},7},
\end{eqnarray}
where $T_{{\rm eff},7}$ is the effective temperature in units of $10^7~{\rm K}$. At ages $t\gtrsim10^5~{\rm yrs}$, $L_{\gamma}$ is affected by the NS surface compositions. For young NSs, the surface composition significantly affects the relation between the surface and interior temperatures at $\rho \simeq 10^{10}~{\rm g~cm^{-3}}$ (often called {\it $T_s$--$T_b$ relation}). In order to treat the uncertainties of the surface composition, we adopt the {\it $T_s$--$T_b$ relation} of \cite{1997A&A...323..415P}, which introduces a parameter:
\begin{eqnarray}
\eta_{\rm PCY97}\equiv g^2_{s,14}\frac{M_{\rm env}}{M_{\rm NS}}
\label{eq:4}
\end{eqnarray}
where $M_{\rm e}$ is the envelope mass and $g_{s,14}$ is the surface gravity in units of $10^{14}~{\rm cm~s^{-2}}$.  If the NS surface is composed of heavy elements such as Fe, then $\eta_{\rm PCY97}=0$. The upper limit of $\eta_{\rm PCY97}$ is almost exclusively determined by the strongly degenerated electron pressure proportional to Eq.~(\ref{eq:4}) for the high-density regime due to the hydrostatic equilibrium condition~(e.g., \citealt{1983ApJ...272..286G}). 
Since the critical pressure corresponding to the maximum value of $\eta_{\rm PCY97}$ depends on the Fermi-wave number density of electrons~(e.g, \citealt{1983bhwd.book.....S}), if the density gets closer to the neutron drip line, it is harder for ions to survive. From such a condition, one can get $\eta_{\rm PCY97}\lesssim10^{-7}$ in case that the matter is fully accreted onto the NS.  The envelope model, as well as the crust superfluid model, is a crucial factor for the thermal luminosity of NS 1987A~\citep{2020ApJ...898..125P}.

In the envelope model of \cite{1997A&A...323..415P}, two cases of pure Fe layer and successive H-He-C-Fe layer were considered, fixing partially-ionized plasma EOSs~\citep{1995ApJS...99..713S,1996ApJ...456..902R}. In latter case, they assumed each density of the interfaces between different elements without their mixture, which could affect the $T_s$--$T_b$ relation. Sophisticated envelope models have been developed to unfix their critical densities including matter mixture, but we note that their uncertainties are not so important for the $T_s$--$T_b$ relation under quasi-stationary envelope evolution~(\citealt{2021PhR...919....1B} and reference therein).

The temperature evolution of NSs is determined by the relativistic heat balance and heat flux equations with a NS structure, obtained by the Tolman-Oppenheimer-Volkoff (TOV) equation. For the numerical calculation of cooling curves, we use the public code \texttt{NSCool}~\citep{2016ascl} with some minor modifications~\citep{2021PTEP.2021i3E01D}. The initial redshifted temperature is fixed to be $10^{10}~{\rm K}$ isothermally. Note that the cooling curves do not depend on the initial temperature for $t\gtrsim10^{-3}~{\rm yrs}$~\citep{2020ApJ...888...97B}.

\section{Comparison between cooling models and observations}
\label{sec:main}

\begin{figure}[t]
    \centering
    \includegraphics[width=0.8\linewidth]{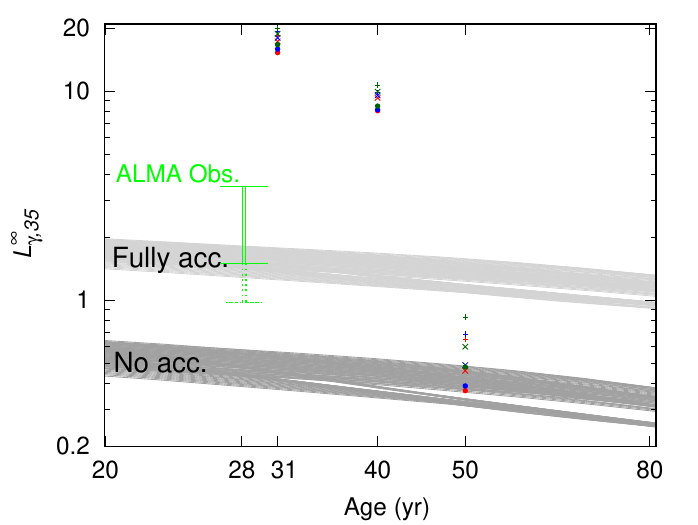}
    \caption{Comparison of cooling curves with ALMA observational constraints of SN 1987A and X-ray limits. The ranges of cooling model parameters in this figure are $\eta_{\rm PCY97}=10^{-6.6}$ (fully-accreted case) and $10^{-15.0}$ (no-accretion case), $M_{\rm NS}=1.18$--$1.62~M_{\odot}$ with the step size of $0.02~M_{\odot}$, and 5 kinds of crust superfluid models: ALL Unpaired ($T_{\rm cr}=0$), CCDK, WAP, T, and ALL ($T_{\rm cr}=10^{10}~{\rm K}$). The green error bars indicate the estimated luminosities from ALMA observation of SN 1987A ($40$--$90~L_{\odot}$) with the downward extension in case that the external heating accounts for 33\% of the observed blob luminosity, i.e. the lower limit from the CCO is $26~L_{\odot}$~\citep{2019ApJ...886...51C,2020ApJ...898..125P}. Symbols denote the upper redshifted luminosities at $t=31,40$ and $50~{\rm yrs}$, as in Fig.~\ref{fig:upper_limits_90_1Ms}.}
    \label{fig:obs}
    \end{figure}
    
    \begin{figure}[t]
    \centering\vspace*{-0.50cm}
    \includegraphics[width=\linewidth]{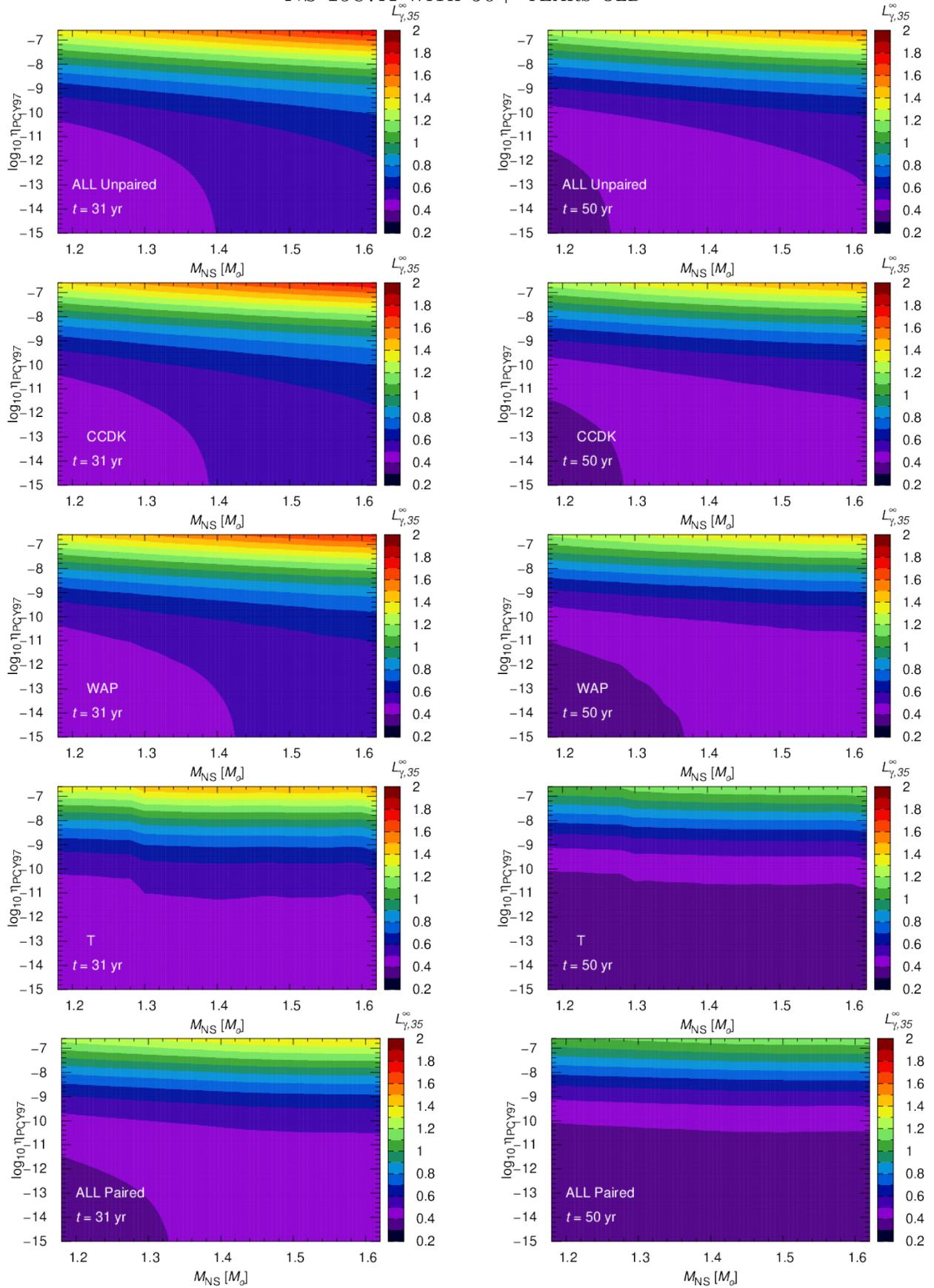}
     \caption{The values of the redshifted bolometric luminosity $L^{\infty}_{\gamma,35}$ (in units of $10^{35}~{\rm erg~s^{-1}}$) in $M_{\rm NS}$--$\eta_{\rm PCY97}$ plane. The age of NSs is $t=31~{\rm yrs}$ for left panels while $t=50~{\rm yrs}$ for right panels. From top to bottom panels, crust superfluid models are All Unpaired, CCDK, WAP, T, and All Paired, respectively.}
     \label{fig:L35}
\end{figure}

    \begin{figure}[t]
    \centering
    \includegraphics[width=0.9\linewidth]{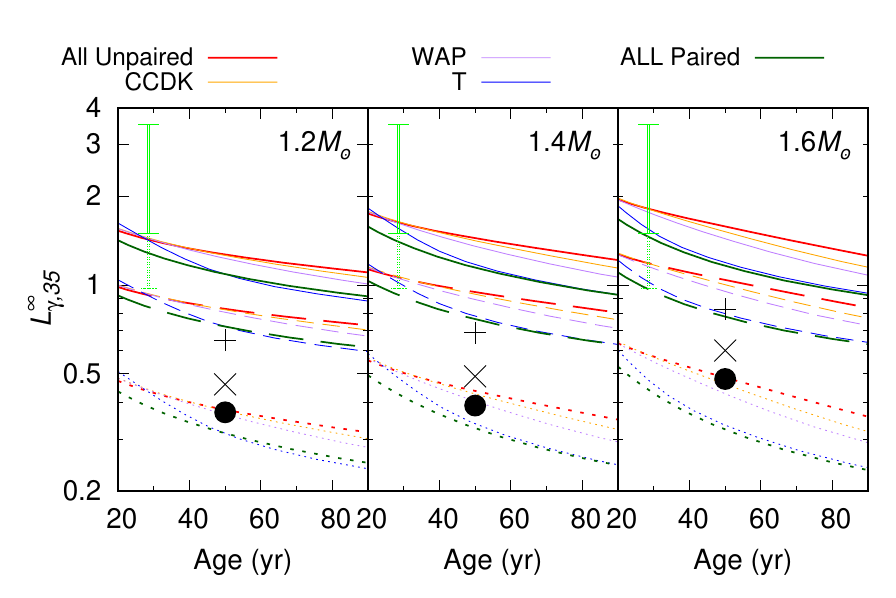}
    \caption{Same cooling curves as Fig.~\ref{fig:obs} but limited at $1.2,1.4$ and $1.6~M_{\odot}$ stars. $\eta_{\rm PCY}=10^{-6.6}$ for solid curves, $\eta_{\rm PCY}=10^{-8}$ for dashed curves and $\eta_{\rm PCY}=10^{-15}$ for dotted curves. As shown in Fig.~\ref{fig:obs}, the circle, cross and plus symbols indicate the bolometric upper limits in the 2040s discussed in Section \ref{sec:obs}, assuming kick velocity of 300 ${\rm km~s^{-1}}$, 500 ${\rm km~s^{-1}}$, and 700 ${\rm km~s^{-1}}$, respectively.}
    \label{fig:exa}
\end{figure}

First, we qualitatively compare the theoretical luminosity values with the X-ray limits\footnote{We note that the luminosities present in Fig.~\ref{fig:upper_limits_90_1Ms} are intrinsic, bolometric and redshifted luminosities; we refer to them as ``X-ray" luminosities to specify that they are derived from X-ray data analysis as seen in Sect.~\ref{sec:obs}. Furthermore, we note that these luminosities are considered as upper limits in Sect.~\ref{sec:obs} under the assumption of non-detection of the NS in SN 1987A. However, as we will discuss later, there are cases for which we predict a possible detection of the NS by \lynx. In this latter cases, the ``X-ray" luminosities are ``lower" limits to the NS detection. In the paper, we refer to these (upper/lower) sensitivity limits as ``X-ray limits".} presented in Fig.~\ref{fig:upper_limits_90_1Ms}. In Fig.~\ref{fig:obs}, we show the comparison between NS cooling models and observations. As we see, the X-ray limits at $t=31$ and $40~{\rm yrs}$ do not allow to constrain any cooling model, even considering higher exposure time (1 Ms) and lower confidence level (1.6$\sigma$). On the other hand, the X-ray limits at $t=50~{\rm yrs}$ challenge some cooling models. Therefore, we focus only on the \lynx-based upper limits in the 2040s.

The ALMA observation should also be considered to constrain the cooling models. Assuming a dust model (``ACAR" sample in \citealt{1996MNRAS.282.1321Z}), \cite{2019ApJ...886...51C} estimated the flux density of the blob detected in SN 1987A and obtained a bolometric luminosity of $(40$--$90)~L_{\odot}$ for 679 GHz flux densities of 1--2 mJy. This {\it estimated} luminosity does not directly represent the luminosity of the CCO, including uncertainties in the flux density measurements and temperature estimate. If there are additional heating effects, the luminosity of the CCO must be lower. For example, it is discussed in \cite{2020ApJ...898..125P} that about 33\% of the luminosity of the blob at 22K may be explained by the heating of ${}^{44}{\rm Ti}$ decays\footnote{From observed line flux at 67.87 keV, the total mass of ${}^{44}{\rm Ti}$ in the ejecta of SN1987A was estimated as $\sim1.5\times10^{-4}~M_{\odot}$, where the optical depth of the ejecta for the line is evaluated as $\sim$ a few $\%$~\citep{2021ApJ...916...76A}. The estimation of the optical depth for hard X-ray/gamma-ray lines from decays of ${}^{44}{\rm Ti}$ is also about $\sim 5\%$~\citep{2011A&A...530A..45J,2015Sci...348..670B}. Thus, the assumed absorption of 33$\%$ of hard X-ray/gamma-ray lines from decays of ${}^{44}{\rm Ti}$ might be overestimation unless other heating mechanisms work significantly.}. Hence, it is worth to mention that the bolometric luminosity estimated for the blob is not a direct observation of the NS and it also reflects some uncertainties derived from the dust model and the contribution of external heating, which denotes heating components by external sources apart from NSs on the thermal emission scenario (``external" heating hereafter). Moreover, if the main heating component comes from non-thermal emission, the ALMA observations can only give upper limits to the thermal luminosity and our conclusions would be significantly different from that of the thermal emission scenario. Nevertheless, we assume that the CCO is responsible for the heating of the blob as mentioned earlier, and we simply take thermal luminosities either with or without the heating corresponding to 22 K dust as the ALMA observation, whose treatment is the same as \cite{2020ApJ...898..125P}. As we see in Fig.~\ref{fig:obs}, the ALMA observation indicates high thermal luminosity in cooling theory, which implies large amounts of accreted matter onto the NS 1987A. Thus, the lower limit of the ALMA observation, including how the external heating contributed to the blob regions, is an important constraint\footnote{We assume that the main heating component is thermal emission emitted on the NS surface. Hence, the external heating considered here does not include the synchrotron heating from PWN~\citep{2021ApJ...908L..45G,gmo22}; this is completely another scenario.}.

The surface luminosity of young NSs highly depends on the NS mass, envelope mass, and crust superfluid models. In Fig.~\ref{fig:L35}, we show the luminosity values at $t=31$ and 50 yr. We cover the parameter regions of $-15.0\le\log_{10}\left(\eta_{\rm PCY97}\right)\le-6.6$ and $1.18\le M_{\rm NS}/M_{\odot}\le1.62$, and their parameter grid intervals are $\Delta\log_{10}\left(\eta_{\rm PCY97}\right)=0.2$ and $\Delta M_{\rm NS}=0.02~M_{\odot}$, respectively. We again note that the considered mass range matches the light-curve observations of SN 1987A~\citep{1988ApJ...330..218W,1988A&A...196..141S,2019A&A...624A.116U,2020ApJ...890...51E,2020ApJ...898..125P}.

The overall surface luminosity at $t=31~{\rm yrs}$ is higher than that at $t=50~{\rm yrs}$ by (0.2--0.4)$\times10^{35}~{\rm erg~s^{-1}}$. If the envelope mass is higher, the luminosity is higher and this trend is valid irrespective of other parameters. If the crust superfluidity is strong, the luminosity is lower at both ages. The physical reasons derive from the enhanced cooling processes of the PBF and the reduction of neutrons specific heat due to their pairing as we describe in Section~\ref{sec:setup}, both of which increase the cooling rate of NSs. The trend of crust superfluidity can be clearly seen in $M_{\rm NS}$--$\eta_{\rm PCY 97}$ planes; If the crust superfluidity becomes stronger, the regions with $L^{\infty}_{\gamma,35}<0.5$ (light purple) for $t=31~{\rm yrs}$ and $L^{\infty}_{\gamma,35}<0.4$ (dark purple) for $t=50~{\rm yrs}$ become wider toward higher-mass regions, respectively.

The X-ray limits in 2027 (2037) are valid only if \chandra\ (\lynx) will not detect thermal X-rays of NS 1987A before 2027 (2040s). It is also noted that \lynx\ will not be launched until the 2040s (see section 4.3).
Then, if the luminosity of NS 1987A is higher than the X-ray limits at the year, NS 1987A should be detected. As we see in Fig.~\ref{fig:exa}, sensitivity limits for a 1 Ms observation from the \lynx\ spectra challenge some theoretical cooling models, which should take the luminosity values between non-accreted matter ($\eta_{\rm PCY97}=10^{-15}$) and fully-accreted matter ($\eta_{\rm PCY97}=10^{-6.6}$) cases. Thus, we can consider two possible scenarios, that is, a non-detection scenario and a detection scenario of the CCO by \lynx. Considering that the \lynx\ is going to be launched in 2036~\citep{2018arXiv180909642T} (see section G-3. and Table G-3. in this paper), since the sensitivity in \chandra\ is not good compared with \lynx\ as we see in Fig.~\ref{fig:obs}\footnote{The reason of the difference of the sensitivity steams from effective area; At $T=1~{\rm keV}$, it is $2~{\rm m}^2$ for X-ray telescope in \lynx\, while $0.08~{\rm m}^2$ for \chandra\,. The on-axis angular resolution is almost the same $\sim0.5$ arcsec.}, the CCO in SN 1987A is unlikely to be detected until 2036\footnote{There are other X-ray satellites which will be launched in near future, but they all have lower spatial resolution than \chandra\,, such as XRISM ($\sim1.2$ arcmin) and Athena ($\sim$5 arcsec).}. Hence, discussing both the detection and non-detection of the CCO by \lynx seems worthwhile. We discuss the two scenarios also taking into account the constraints by the ALMA observation in subsections \ref{subsec:4.1} and \ref{subsec:4.2}, respectively.

 \subsection{Non-Detection Case of NS 1987A}
\label{subsec:4.1}
If the thermal emission from NS 1987A will not be detected by \lynx, the NS cooling models could be constrained by the \lynx\ X-ray limits. In Fig.~\ref{fig:exa}, we compare the cooling models with the ALMA observation and the \lynx\ upper limits. We choose three envelope models of $\eta_{\rm PCY97}=10^{-6.6},10^{-8},10^{-15}$ and NS masses of $M_{\rm NS}=1.2,1.4,1.6~M_{\odot}$, respectively. NS cooling models with the highest envelope mass of $\eta_{\rm PCY97}=10^{-6.6}$ are always compatible with the ALMA observation. Meanwhile, \lynx\ upper limits prefer the lower envelope models of $\eta_{\rm PCY97}\sim10^{-15}$, whose luminosity is almost the same with the non-accreted matter models. Thus, the two extreme models for the envelope mass cannot be compatible with both observations simultaneously, although if the contribution for external heating in  ALMA observations is even higher, the cooling curves might be consistent with the ALMA observation even considering low-mass envelope. In the fine-tuned moderate envelope masses, both observations could be explained by choosing strong crust superfluid models. This is because the observational luminosity inferred by ALMA at 28 yrs is high while \lynx\ upper limits are low compared with ALMA, which implies that rapid cooling triggered by crust superfluidity around $t\sim40~{\rm yrs}$ is required. For example, the $1.6~M_{\odot}$-stars model with $\eta_{\rm PCY97}\lesssim10^{-8}$ and the crust superfluid model of T can account for the ALMA observation and the \lynx\ upper limits with $v_{\rm kick}=700~{\rm km~s^{-1}}$. Thus we find that the two following requirements must be satisfied at the same time to account for the independent observations\footnote{If the external heating to the observed blob is more significant than considered here (33\%), the lower $v_{kick}$ and envelope-mass values can be allowed. However, we assume the thermal emission scenario and the external heating as a representative of ${}^{44}{\rm Ti}$ may not work so much as in the footnote 4, ALMA observations must have at least lower limits of luminosity, and therefore the qualitative constraints are expected to be unchanged.}:
\begin{itemize}
    \item Accreted matter must exist on NS 1987A, but the envelope mass is severely constrained. 
    \item A rapid cooling must have occurred around $t\simeq40~{\rm yrs}$. 
\end{itemize}
The first remark implies that most cooling models would be rejected with a non-detection, except $\eta_{\rm PCY}\simeq10^{-8}$ for the APR EOS. The second remark implies that the PBF process powerfully works in the crust, and thus, the crust superfluidity must be vital in adjusted $\eta_{\rm PCY97}$ values. This means that the future X-ray observations by \lynx\ is rather important to test NS cooling models. On the other hand, for $M_{\rm NS}=1.2,1.4M_{\odot}$ stars, there is no parameter regions able to explain both observations. There are two reasons for this. 
One is that \lynx\ upper limits are higher with a higher mass due to the higher redshift of blackbody radiation, as mentioned in Section~\ref{sec:obs}. The other one is the difference between $g_{s,14}$ values, which slightly change the initial relaxation time scale~(see Figure 9 and compare between Model A and C in \citealt{2020ApJ...888...97B}). According to that, the theoretical luminosity with $t=10^{0-2}~{\rm yrs}$ tends to be higher for higher $g_{s,14}$ values. Although the EOS uncertainties (including the heat capacity and the conductive opacity) may change the consistency, the NS mass could also be constrained depending on the envelope and crust superfluid models.

  \begin{figure}[t]
    \centering
    \includegraphics[width=\linewidth]{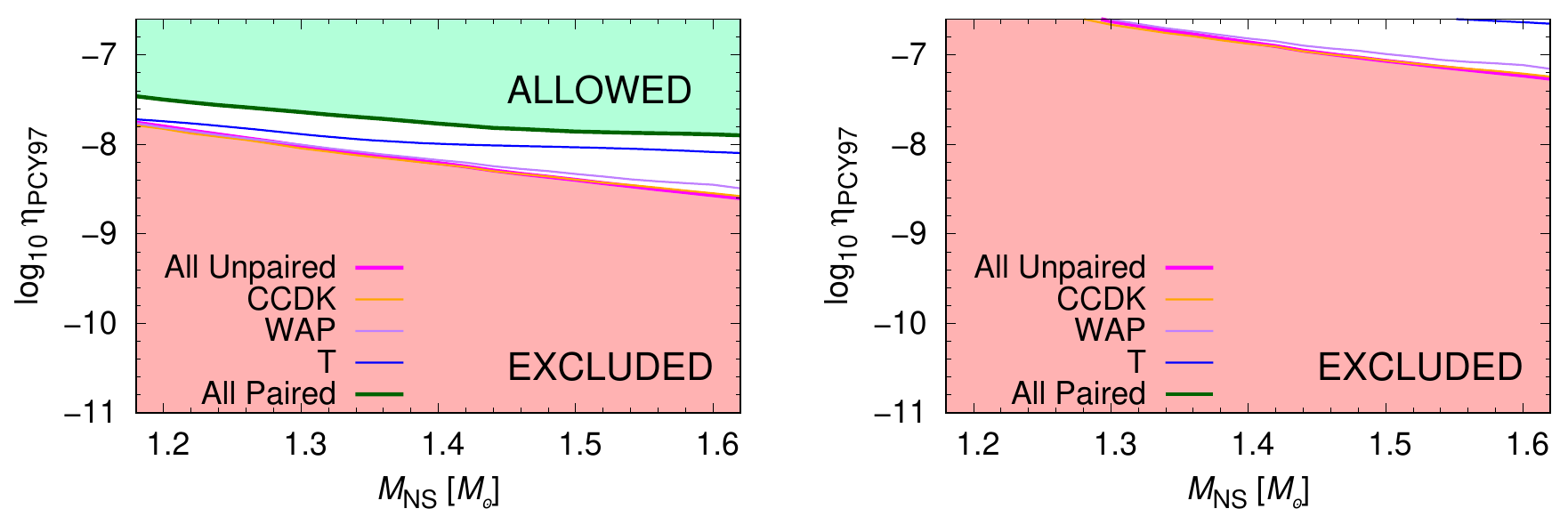}
     \caption{Constraints on cooling models (i.e., NS mass, envelope mass, and crust superfluidity) of NS 1987A from ALMA observation, where the lower luminosity value is $26~L_{\odot}$ (left) and $40~L_{\odot}$ (right), corresponding to the percentage of the external heating in the observed blob being 33\% and 0\%, respectively. Whether non-colored regions are allowed or excluded depends on the adopted crust superfluid model.}
    \label{fig:lower1}
\end{figure}

Using various cooling models, we investigate the allowed parameter regions for the ALMA lower limits and the \lynx\ upper limits by comparing the theoretical luminosity with the observational one in each grid. First, we show the ALMA observational constraints on cooling models of NS 1987A in Fig.~\ref{fig:lower1}.

As reported by \cite{2020ApJ...898..125P}, the ALMA observation rejects many cooling models except high-mass envelopes, and weaker crust superfluidity is preferred in some envelope masses. For higher NS masses, the allowed regions become wider because the theoretical luminosity value is higher for higher mass. These trends are valid regardless of how the external heating, except thermal emission, contributes to the observed blob luminosity. If the external heating does not contribute to the observed blob in SN 1987A, the NS mass could be automatically constrained for some crust superfluid models because of $\eta_{\rm PCY97}\lesssim10^{-7}$; For example, we can give $M_{\rm NS}>1.4~M_{\odot}$ for the strong crust superfluid model of All Paired. 

  \begin{figure}[t]
    \centering
    \includegraphics[width=\linewidth]{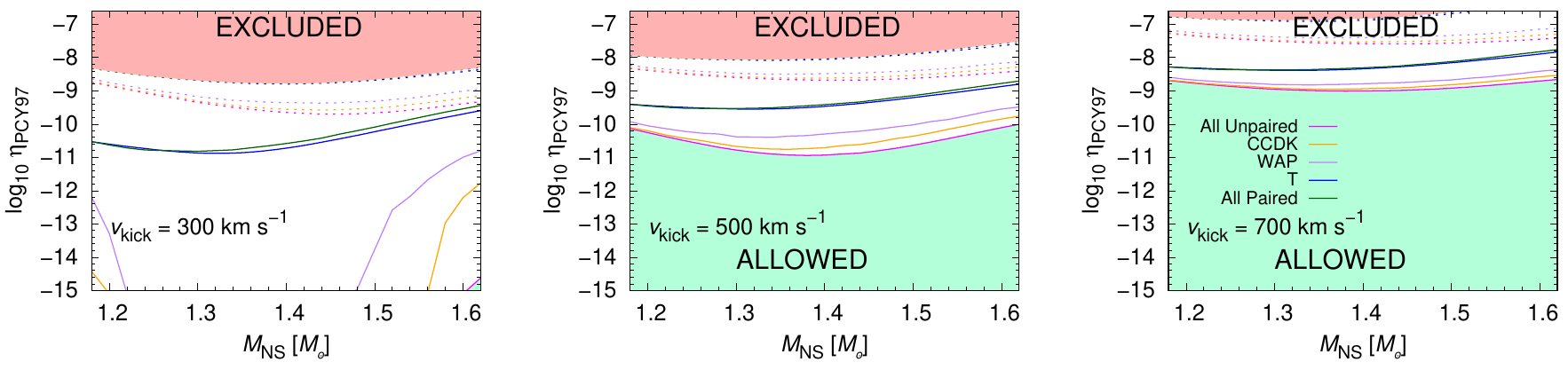}
    \caption{Constraints on cooling models from the \lynx\ spectra with an exposure time of 1 Ms ($2\sigma$ errors for solid curves while $3\sigma$ for dotted). Three cases of the NS kick velocities, $v_{\rm kick}=300~{\rm km~s^{-1}}$(left), $500~{\rm km~s^{-1}}$(middle), and $700~{\rm km~s^{-1}}$(right), are shown. Excluded regions are shown in red color, allowed in green. The non-colored area reflects the uncertainty in the superfluidity model, and might be allowed or excluded depending on the assumed one.}
    \label{fig:upper1}
\end{figure}

Second, we show the constraints by \lynx\ upper limits in Fig.~\ref{fig:upper1}. For the \lynx\ upper limits except for $M_{\rm NS}=1.2,1.4$ and $1.6~M_{\odot}$, we take a quadratic interpolation  among them. The most crucial parameter for constraints on the cooling models is the NS kick velocity, $v_{\rm kick}$. As mentioned in Section~\ref{sec:obs}, the X-ray limits of higher-$v_{\rm kick}$ values become higher, which allows more cooling models. This can be clearly seen in Fig.~\ref{fig:upper1}; For $v_{\rm kick}=300~{\rm km~s^{-1}}$ with 2$\sigma$ errors, the NS mass could be constrained with weak crust superfluid models. To satisfy both constraints by ALMA and \lynx\,, the most important parameter for the cooling models is the envelope mass, which must be small. The second one is the crust superfluidity and the third is the NS mass. 

 \begin{figure}[t]
    \centering
    \includegraphics[width=\linewidth]{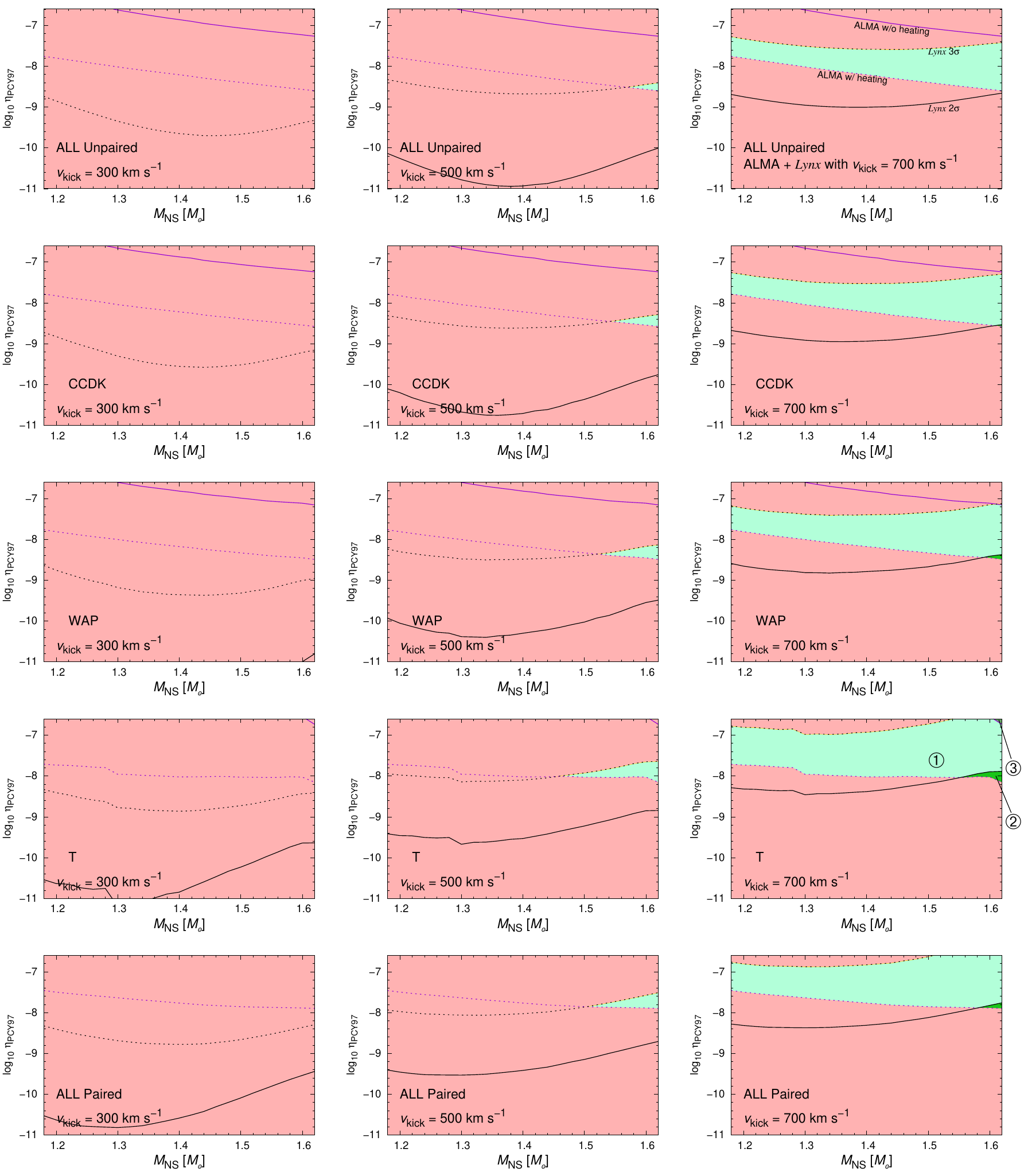}
    \caption{Constraints on NS 1987A models by the ALMA observation (purple lines: solid and dotted lines are without and with external heating, respectively) and \lynx\, upper limits (black lines: solid and dotted lines are for 2$\sigma$ and 3$\sigma$ errors, respectively) in the 2040s. 15 figures are shown for different $v_{\rm kick}$ and crust superfluid models separately. The different green colors (\textcircled{\scriptsize 1}--\textcircled{\scriptsize 3}) indicate the regions which satisfy the following observations; green (\textcircled{\scriptsize 1}) for the ALMA with an external heating of one-third of the blob luminosity and for the \lynx\ with $3\sigma$ errors, web-green (\textcircled{\scriptsize 2}) for ALMA with the external heating and \lynx\ with $2\sigma$ errors, and dark-green (\textcircled{\scriptsize 3}) for ALMA without the external heating and the \lynx\ with $3\sigma$ errors. There are no allowed regions which satisfy both the ALMA without the external heating and the \lynx\ with $2\sigma$ errors.}
    \label{fig:undetected}
\end{figure}

 In Fig.~\ref{fig:undetected}, we show the parameter regions constrained by the ALMA and \lynx\ observations. Remarkably, there is no allowed region for the case of $v_{\rm kick}=300~{\rm km~s^{-1}}$ because the \lynx\, upper limits are lower than those with other $v_{\rm kick}$. For $v_{\rm kick}=700~{\rm km~s^{-1}}$, on the other hand, there are allowed regions regardless of the NS mass, considering the $3\sigma$ errors of \lynx\ upper limits. We note that, if the percentage of external heating is higher than 33\%, lower $v_{\rm kick}\sim300~{\rm km~s^{-1}}$ is quite possible (for instance, suppose the radioactive heating to 30 K by some external sources contributes to the observed blob luminosity, then the lower thermal luminosity is reduced to 32\% of the blob luminosity ($L^{\infty}_{\gamma,35}=0.48$), which is compatible with \lynx\, upper limits with $v_{\rm kick}=300~{\rm km~s^{-1}}$, as we see Fig.~\ref{fig:exa}.  For $v_{\rm kick}=500~{\rm km~s^{-1}}$, only masses larger than $M_{\rm NS}\gtrsim1.5~M_{\odot}$ are allowed. Thus, our results show that the kick velocity is likely to be as high as $\sim700~{\rm km~s^{-1}}$ for the non-detection scenario. Thus, our results show that the kick velocity is likely to be as high as 700 ${\rm km~s^{-1}}$, provided that the NS 1987A is not found by \lynx\, observations, i.e., non-detection scenario. In other words, future observations of $v_{\rm kick}$ are quite important for judging the detectability by \lynx\,. For instance, suppose that they give low $v_{\rm kick}$ (~$300~{\rm km~s^{-1}}$), then the NS1987A is likely be detected in the 2040s. Although the NS kick velocity is the most crucial parameter, other cooling model parameters could also be constrained by ALMA and \lynx\,. For both cases with $v_{\rm kick}=500$ and $700~{\rm km~s^{-1}}$, the envelope masses are likely to be higher with stronger crust superfluid models as we see in Fig.~\ref{fig:undetected}. Thus, the crust superfluidity also affects the allowed cooling model parameters. If $v_{\rm kick}=500~{\rm km~s^{-1}}$, the lower limit of the NS mass could be changed in the range from $M_{\rm NS}\simeq1.48~M_{\odot}$ to $1.56~M_{\odot}$. If $v_{\rm kick}=700~{\rm km~s^{-1}}$, some strong crust superfluid models with high NS mass regions, i.e.  web-green (dark-green) regions, could also be allowed by the \lynx\, upper limits with $2\sigma$ errors and the ALMA observations with external heating (the \lynx\, upper limits with $3\sigma$ errors and the ALMA observations without external heating). Then, the mass of NS 1987A is almost determined to be $M_{\rm NS}\sim1.6~M_{\odot}$ from the current prediction of 1.22 $\le M_{\rm NS}/M_{\odot}\le$ 1.62~\citep{2019A&A...624A.116U,2020ApJ...890...51E,2020ApJ...898..125P}.
 
The above discussion is based on the assumption that the CCO is responsible for the heating of the blob in SN 1987A. Otherwise, i.e., in the non-thermal emission scenario, the ALMA observations can be regarded as upper limits for the thermal luminosity; Then, the allowed/excluded regions from ALMA observational constraints in Fig.~\ref{fig:lower1} become opposite and most of cooling models except with light envelopes could be allowed. Since both ALMA observations and X-ray limits have \underline{upper} limits in common, the \lynx\, X-ray limits could reject a few cooling models. In non-thermal emission scenario, therefore, we cannot strongly mention the possibility that the CCO will be found by \lynx\,. Nevertheless, thermal emission from the NS is the most plausible for explaining the blob luminosity in a number of scenarios~(see Table 1 in \cite{2020ApJ...898..125P}). Thus, we can conclude from Fig.~\ref{fig:undetected} that, if the CCO is likely not to be found by \lynx\,, the CCO would not contribute to the ALMA observation.
 
 \subsection{Detection Case of NS 1987A}
 \label{subsec:4.2}
 
  \begin{figure}[t]
    \centering
    \includegraphics[width=\linewidth]{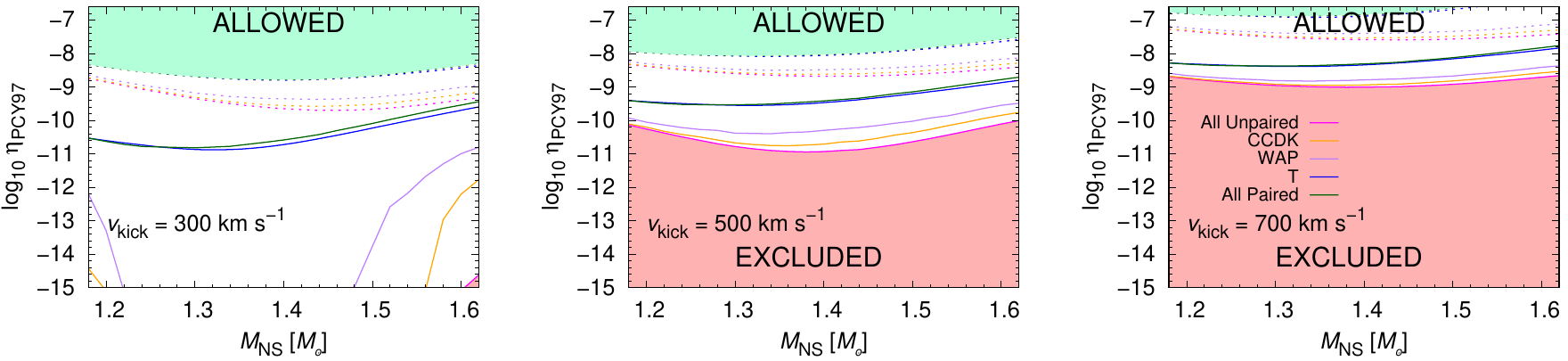}
    \caption{Same as Fig.~\ref{fig:upper1}, but in case of a future detection of NS 1987A with \lynx.}
     \label{fig:upper2}
    \end{figure}
    
  \begin{figure}[t]
    \centering
      \includegraphics[width=\linewidth]{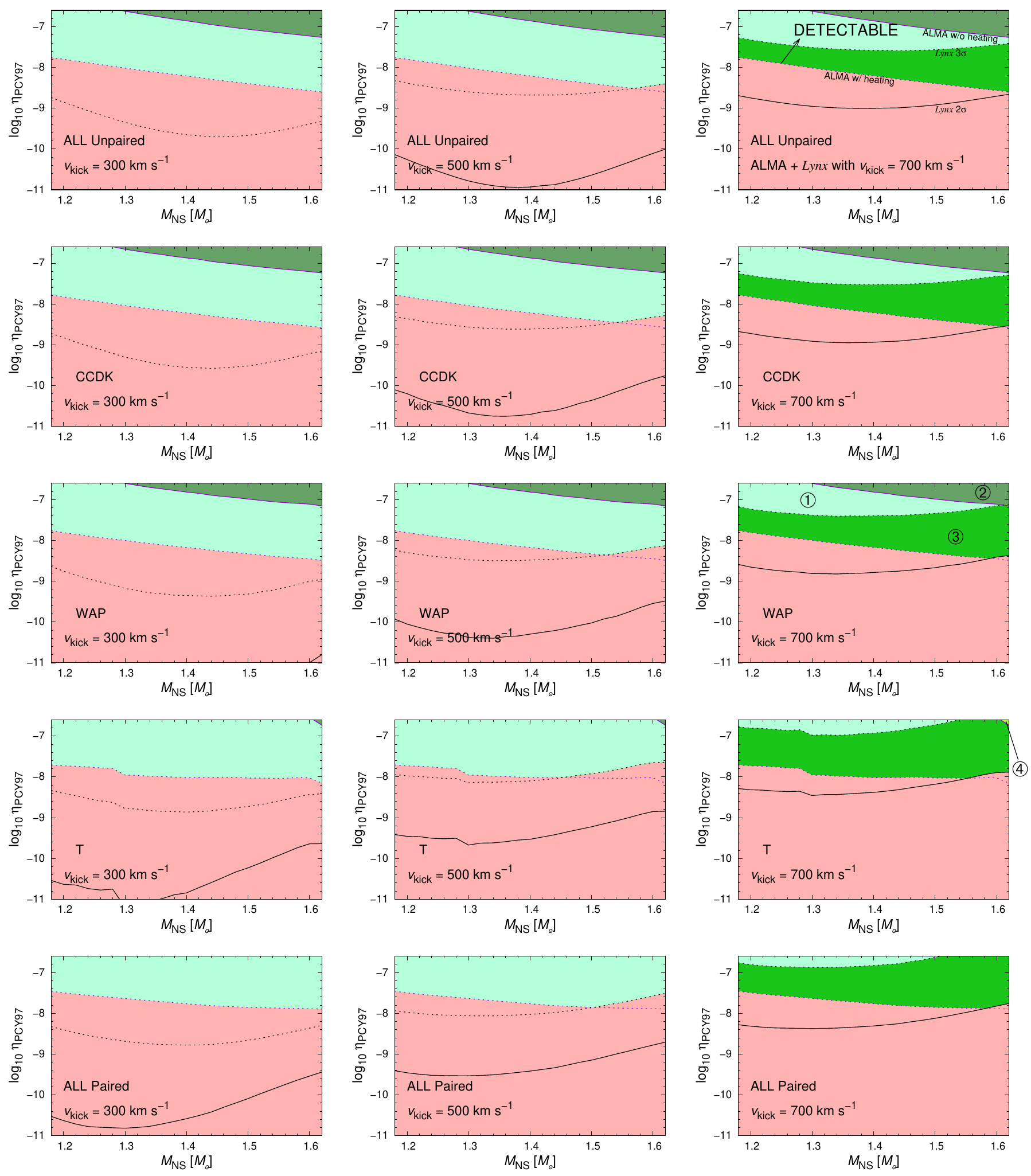}
    \caption{Constraints on NS 1987A models by the ALMA observations and the \lynx\, lower limits for a possible detection of NS 1987A with $t\gtrsim50$ years old. The lines, adopted models and observations including their uncertainties (color) for each panel are the same as Fig.~\ref{fig:undetected}, but we add the pea-green (\textcircled{\scriptsize 4}) regions of the observationally possible detection satisfying the ALMA without the external heating and \lynx\ with $2\sigma$ errors. }
    \label{fig:detected}
\end{figure}

\begin{figure}[t]
    \centering
     \includegraphics[width=0.8\linewidth]{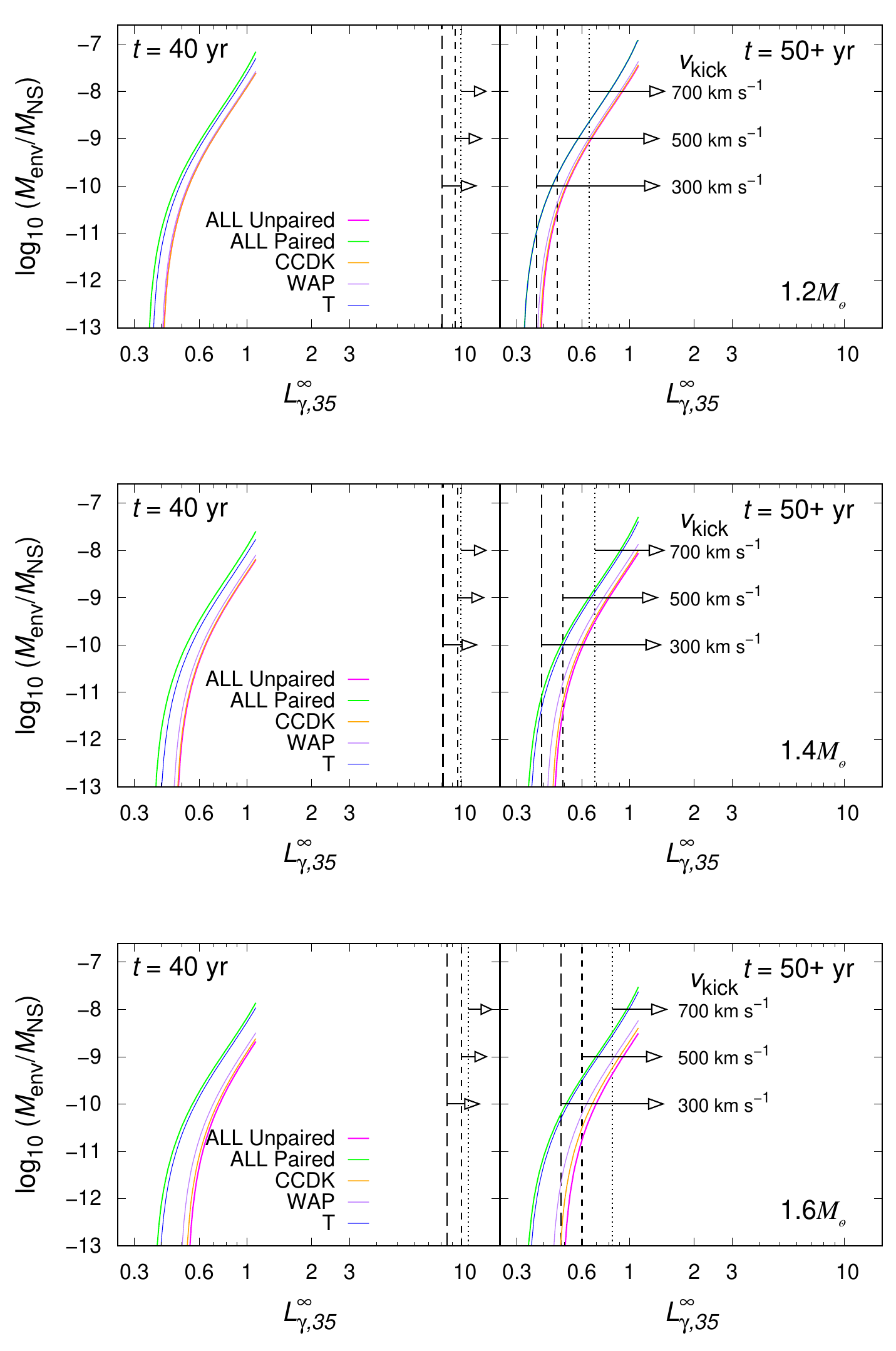}
    \caption{Cooling models (solid lines) in the plane of the envelope mass and the redshifted luminosity. The lower redshifted luminosity limits for a detection of NS 1987A by \chandra\, (left: $t$ = 40 yrs) and \lynx\, (right: $t$ = 50 yrs) are also shown for the NS kick velocities of 300 km s$^{-1}$ (long dashed line), 500 km s$^{-1}$ (dashed line), and 700 km s$^{-1}$ (dotted line) with 2$\sigma$ errors. From top to bottom panels, the NS masses are 1.2$M_{\odot}$, 1.4$M_{\odot}$, and 1.6$M_{\odot}$, respectively.}
    \label{fig:pre}
\end{figure}

Now suppose that black-body radiation in SN 1987A is observed by \lynx\, and thus NS 1987A is confirmed to exist, i.e., the flux derived from theoretical luminosity is higher than the threshold flux obtained from that luminosity for the detection of NS 1987A with \lynx\, as shown in Fig.~4. The parameter regions allowed for this case are shown in Fig.~\ref{fig:upper2}. The higher the NS kick velocity is, the higher the envelope mass is. Since \lynx\, data give lower luminosity limits for a detection of NS 1987A, a high envelope mass is favored, which is consistent with constraints on the envelope mass by ALMA observations as seen in Fig.~\ref{fig:exa}.

Then, we also consider the ALMA constraints. The combined parameter regions of detectability are shown in Fig.~\ref{fig:detected}. For $v_{\rm kick}=300~{\rm km~s^{-1}}$, the allowed regions are determined only by the ALMA observation. For $v_{\rm kick}=500$ and $700~{\rm km~s^{-1}}$, on the other hand, some allowed regions for high NS masses could be slightly reduced compared with the original ALMA constraints, which indicates higher envelope masses. In particular, the allowed regions depend on the observational uncertainties of \lynx\, for $v_{\rm kick}=700~{\rm km~s^{-1}}$ cases. For example, if we  consider the $3\sigma$ errors for the \lynx\ observations high-envelope mass and high NS mass are favored.  

In Fig.~\ref{fig:pre}, we show how the model parameters of NS 1987A can be constrained, if NS 1987A with $t\gtrsim50$ years old will be detected by \lynx\,\footnote{Only here, we adopt $M_{\rm env}$ not $\eta_{\rm PCY97}$ because $g_{s,14}$ value is constant for each panel with 1.2, 1.4, and 1.6 $M_{\odot}$ stars, respectively. Furthermore, the envelope mass is more intuitively clear than $\eta_{\rm PCY97}$ and could be directly observed. Actually, the envelope mass is much more used in various fields related to SN 1987A than $\eta_{\rm PCY97}$, except in the NS cooling theory.}. Here, the ALMA constraints are not considered. At $t=40~{\rm yrs}$ (2027), the X-ray limits are derived from synthetic \chandra\ spectra and are around $10^{36}~{\rm erg~s^{-1}}$, which is higher roughly by an order of magnitude than that for thermal emission from an NS at the 2040s. Therefore, there is little chance to identify the CCO by detecting the thermal X-ray emission directly in 2027.

On the other hand, the upper limits by \lynx are comparable to theoretically calculated thermal luminosities, so there is some chance to detect the CCO. Depending on the observed redshifted luminosity $L_{\gamma}^{\infty}$ for $t\gtrsim50~{\rm yrs}$, the NS kick velocity could be constrained. Let us assume $M_{\rm NS}=1.6~M_{\odot}$ and $L_{\gamma}^{\infty}$ ($t=50~{\rm yrs}$) $= 5\times10^{34}~{\rm erg~s^{-1}}$, then the NS kick velocity can be speculated as $v_{\rm kick}<500~{\rm km~s^{-1}}$, which is consistent with $v_{\rm kick}\lesssim700~{\rm km~s^{-1}}$ from ALMA observations. Thus, we can conclude that the NS in SN 1987A could be detected by \lynx\ around 2040 as long as the thermal emission scenario is correct.

\subsection{On the Delay of \lynx\ Being Launched}

\begin{table}[h]
    \centering
    \caption{Some X-ray luminosity limits for detectability of NS 1987A with 1 Ms exposure time.}
    \begin{tabular}{c|ccc}
    \hline\hline
    $(M_{\rm NS},v_{\rm kick})$ & Confidence Level & X-ray limits in 2037 ($10^{35}~{\rm erg~s^{-1}}$) & X-ray limits in 2043 ($10^{35}~{\rm erg~s^{-1}}$)\\
    \hline
     \multirow{2}{*}{$(1.2 M_{\odot}, 300~{\rm km~s^{-1}})$} & 99.7\% & 0.56 & 0.41\\
     & 90\% & 0.37 & 0.28\\
     \hline
     \multirow{2}{*}{$(1.6 M_{\odot}, 700~{\rm km~s^{-1}})$} & 99.7\% & 1.25 & 0.81\\
     & 90\% & 0.83 & 0.55\\
\hline     
    \end{tabular}
    \label{tab:2043}
\end{table}

We discussed the scenario of the NS 1987A being detected in the 2040s, but in reality, took both theoretical and X-ray luminosities in 2037, due to the limitation of the ejecta profile made up to 2037~\citep{oon20}. While 2037 was desired launched year according to the \lynx interim report~\citep{2018arXiv180909642T},
the recent subsequent NAS Decadal Survey did not recommend the development of \lynx\ on such an early timescale~\citep{2021pdaa.book.....N}: The early 2040s seems realistic as the launched year as a result of reasonable assessments of a budget profile, scientific performance, and technology risk\footnote{While the \lynx\ teams proposed presented a 15-year program with a budget of $6.2$ billion, the technical, risk, and cost evaluation analysis showed a 19-year program with a budget of $6.9$ billion. In addition to this, an additional 3+ years are estimated to be required for all missions of \lynx\ according to NAS Decadal Survey.}. Considering the delay in the launch of \lynx\,, we assume 2043 as the launch year, and estimate the X-ray limits at that time in the same manner as the spectral fitting in 2037. In our absorption model of \texttt{TBabs*(vnei+vnei+vphabs(zashift(bbodyrad)))}, we assume that the difference in spectral fitting between 2037 and 2043 is only due to the difference of the column density of the ejecta\footnote{Chemical abundances of the ejecta (along the line of sight) also affect the X-ray absorption, but over a short period of 5 years, they are not changed at all.}. The variation of the column density is determined by the free expansion of the ejecta profile of SN 1987A at 2037~\citep{oon20} connecting to X-ray absorption. Since high $v_{\rm kick}$ and high mass models tend to show higher X-ray fluxes for the detection of \lynx\,, we examine the most extreme scenarios of the detectability of NS 1987A, that is, $(M_{\rm NS},v_{\rm kick})=(1.2 M_{\odot}, 300~{\rm km~s^{-1}})$ and $(1.6 M_{\odot}, 700~{\rm km~s^{-1}})$. The results are shown in Table \ref{tab:2043}. As we can see, the decrease of the X-ray limit for the former scenario is roughly 25\%, while for the latter scenario is 35\%\footnote{Note that these variations are quite similar to that reduction of the ejecta density assuming the free expansion, 28.8\%. Thus, the time evolution of the sparseness of shocked ejecta roughly corresponds to that of the decrease of X-ray limits.}. 

In our minimal NS cooling models, the decrease of theoretical luminosities between 2037 and 2043 results in 2--7\% depending on crust superfluidity, mass, and $\eta_{\rm PCY 97}$. Such a \textit{mild} NS cooling is due to its youth, which implies that the thermal relaxation time has not been reached yet without fast cooling processes. Thus, the \textit{net} decrease of critical luminosity for the detection of NS 1987A between 2037 and 2043 can be estimated to be $\sim20$--$30\%$, which leads to a little higher possibility to detect the NS 1987A by \lynx\ with the passage of years compared to that deduced in this work. Nevertheless, since such a variation does not change the model constraints in 2037 significantly, we can conclude that our results presented so far qualitatively hold in not only 2037 but also since then up to 2043. Namely, it does not matter some delay of \lynx\ observations for our paper. Thus, we again suggest that \lynx\ is the most promising X-ray satellite for the detection of NS 1987A regardless of the possible delay.

\section{Discussions and Concluding Remarks}
\label{sec:conc}

In this paper, we investigated the thermal emission scenario for the hot dust blob in SN 1987A, focusing on the consistency of cooling models with the ALMA and \chandra\, observations and the possible future detection of the CCO by \chandra\ and \lynx\ X-ray observations. We derived the following main conclusions:

\begin{itemize}
\item In all cooling models presented in this paper, the luminosities are lower by orders of magnitudes than the upper limits derived from \chandra\, data, which therefore cannot constrain any model parameters. Hence, according to our results, currently it is almost impossible for \chandra\, to detect the CCO.
    \item The theoretical luminosities are comparable with the upper limits derived for \lynx\, in the 2040s. Namely, several cooling models with high-mass envelopes can be excluded if the CCO will not be detected before the 2040s. The higher the NS kick velocity, the higher the upper limit, and more cooling model parameters are allowed.
    \item The constraints on the model parameters by the \lynx\ upper limits, i.e., the non-detection case, are in contrast with ALMA observational constraints, although this contrast might disappear considering higher percentages of the external heating. Hence, most NS cooling models could potentially be rejected by combining the two independent observations. As a result of the detailed constraints, the NS kick velocity is likely to be as high as $\sim700~{\rm km~s^{-1}}$, assuming that the CCO is responsible for the heating of blob. For the X-ray observations of the high kick velocity, the NS mass is preferred to be as high as $\sim1.6~M_{\odot}$. According to models that satisfy the constraints from both observations, the envelope mass must be as high as $\sim10^{-8}M_{\rm NS}$. At the same time, a rapid neutrino cooling process, which is likely to be PBF triggered by the crust superfluidity, must work at $t\simeq40~{\rm yrs}$.
    \item Since \lynx\, upper limits (at $t=50~{\rm yrs}$) are comparable to the theoretical luminosities, \lynx\ could succeed in the detection of NS 1987A with 50+ years old. The NS must have a high envelope mass in most cases, especially for $v_{\rm kick}=700~{\rm km~s^{-1}}$. Thus, launching the \lynx\ satellite as early as possible is desired for the detection of the youngest NS.
    \item Even with some delay of \lynx\ being launched up to the early 2040s, our results are not so changed compared to those in 2037.
\end{itemize}

In the present study, the assumed kick velocity of NS 1987A is one of key parameters for constraining other NS properties through the estimation of the X-ray absorption and the upper (lower) intrinsic luminosities in case of a non-detection (detection). 
Theoretically, it has been expected that new born NSs are kicked by an asymmetric explosion with an aid by  so-called gravitational tugboat mechanism \citep{2013A&A...552A.126W, 2017ApJ...837...84J} and/or by an anisotropic neutrino emission \citep[e.g.,][]{1987IAUS..125..255W, 2005ApJ...632..531S}. The recently found lepton-number emission self-sustained asymmetry
\citep[LESA:][]{2014ApJ...792...96T} could also contribute to the acceleration of NSs, although the contribution may not be so large \citep{2021ApJ...915...28B}. 
Meanwhile, observed NS kick velocities for young pulsars typically range over several 100 km s$^{-1}$ \citep[e.g.,][]{2006ApJ...643..332F} with some exceptions of over 1000 km s$^{-1}$ \citep{2005ApJ...630L..61C}. Additionally, analyses of X-ray observations of several supernova remnants \citep{2018ApJ...856...18K} have supported a hydrodynamical origin of NS kicks related to anisotropic mass ejection \citep{1996PhRvL..76..352B,2000ApJS..127..141N,2013A&A...552A.126W}. The explosion of SN 1987A has been suggested to have been asymmetric from observed line emission, e.g., iron lines ([Fe II]) \citep{1990ApJ...360..257H} and gamma-ray lines from the decay of $^{44}$Ti \citep{2011ApJS..197...31S}. Actually, the 3D morphology of the inner ejecta of SN 1987A is globally asymmetric with an elongated structure \citep[e.g.,][]{2013ApJ...768...89L, 2016ApJ...833..147L}, which was foreseen in some early numerical simulations of SN 1987A \citep{1997ApJ...486.1026N,1998ApJ...495..413N,2000ApJS..127..141N}. Thus, a kick to NS 1987A has highly been expected from such observations. 
In the 3D MHD model \citep{oon20} used for the estimation of X-ray limits, the NS kick velocity of NS 1987A was estimated as approximately 300 km s$^{-1}$ by assuming momentum conservation. 
Therefore, the range of NS kick velocity values considered in this work (300-700 ${\rm km~s^{-1}}$), which stand in both the detection and non-detection scenario, is roughly consistent with the observed values for young pulsars as well as that suggested from the ALMA observation \citep{2019ApJ...886...51C}. 
In the 3D model, the explosion was, however,  artificially initiated with an ad-hoc asymmetry (asymmetric thermal bomb) \citep{2020ApJ...888..111O}, and the early evolution ($<$ 1 sec) of the shock revival including a fall back was not treated in a realistic way. Although the 3D model \citep{2020ApJ...888..111O, oon20} explains both the observed asymmetric line profiles of iron \citep{1990ApJ...360..257H} and titanium \citep{2015Sci...348..670B} and the X-ray light curves \citep[e.g.,][]{2016ApJ...829...40F} well with a blue supergiant progenitor model resulting from a binary merger \citep{2018MNRAS.473L.101U}, the estimated value may only provide a reference value. 
 Hitherto, recent 2D and 3D more self-consistent neutrino-driven core-collapse supernova simulations 
(e.g., \citealp{2017ApJS..229...42N, 2019ApJ...878..160N, 2021ApJ...915...28B}, for SN 1987A, \citealp{2022arXiv220206295N}) have tried to figure out NS properties. A consensus on the magnitudes of NS kick velocities has, however, not been achieved yet. Therefore, a future determination of the kick velocity of NS 1987A based on the detection by \lynx \ will shed light not only on the NS properties but also on the explosion mechanism and the origin of the explosion asymmetries.

We fix the NS EOS as the APR model in this paper. However, the NS EOS in high-density regions, including the internal compositions, has significant uncertainties even with recent experiments and observations~(for a review, see \citealt{2021PrPNP.12003879B}). A crucial problem for NS cooling theories is the presence or absence of the DU process in the core. Since we limit the NS mass as $1.2\le M_{\rm NS}/M_{\odot}\le1.6$ according to the recent predictions~\citep{2019A&A...624A.116U,2020ApJ...890...51E,2020ApJ...898..125P}, we do not consider the possibility that the DU process occurs for the APR EOS. However, in the case of large-symmetry-energy EOSs, theoretical luminosities should be decreased due to an effective neutrino cooling by the DU process, which could change the consistency with the observations of SN 1987A. Although the recent experiment to measure the neutron skin thickness of $^{208}{\rm Pb}$, the updated lead Radius EXperiment (PREX-2), resulted in the symmetry-energy value of the slope parameter $L=\left(106\pm37\right)~{\rm MeV}$~(\cite{2021PhRvL.126q2503R}, but see also \cite{2022arXiv220603134R}), most experiments and observations show the relatively low symmetry energy of $L\simeq60\pm20~{\rm MeV}$~(e.g., \cite{2022PTEP.2022d1D01S}). Since the $L$ value has a negative correlation with the threshold mass of the DU process~\citep{2019PTEP.2019k3E01D}, the DU process is likely not operating in NS 1987A because of inferred low-mass NSs\footnote{One can get the empirical relation between $L$ and the threshold mass of the DU process $M_{\rm DU}$ \citep{2022PhRvD.105b3007S}: 
\begin{eqnarray}
\frac{M_{\rm DU}}{M_{\odot}} = 3.5801 - 4.2036L_{100}+1.5191L_{100}^2~,
\end{eqnarray}
where $L_{100}\equiv L/\left(100~{\rm MeV}\right)$. Taking the most fiducial value of $L=60~{\rm MeV}$, $M_{\rm DU}\simeq1.6048~M_{\odot}$, which is roughly the upper mass of NS 1987A inferred from observed SN 1987A light curves.
}.

Beyond the nucleon-DU process, other rapid cooling processes through exotic states such as the hyperon DU processes and deconfined quark beta decay may also work and cool young NSs rapidly. However, one of concerns for the rapid cooling scenario for NS 1987A is that the young NS probably has a non-isothermal temperature structure where the crust is hotter than the core. In other words, unless there are large regions where the rapid cooling processes work (in relatively lower-density regions), only the heat in the crust, not in the core, may be transported to the surface, so that the rapid cooling processes may not affect the luminosity around the age of NS 1987A. A candidate as an {\it effective} exotic cooling process which appears in lower-density regions is the pion condensation process, although it can significantly soften the EOS enough not to support the $2~M_{\odot}$ stars~\citep{2022IJMPE..3150006D}. In that sense, the minimal cooling scenario considered here seems reasonable for NS 1987A.

For young NSs around at $t\sim10~{\rm yrs}$, the most crucial cooling process is the PBF process triggered by the crust superfluidity~\citep{2009ApJ...707.1131P} because of the non-isothermal temperature structure as described above. The efficiency of the crust PBF process is determined by the density dependence of the critical temperature, being still uncertain. Since the crust superfluidity can also be probed by the ultra-cold atom gas studies in the context of the BCS-BEC crossover~(for reviews, see \citealt{2019EPJA...55..167S,2020PrPNP.11103739O}), investigating the detailed effects of crust superfluidity from both astrophysics and condensed-matter physics is worthwhile for understanding the properties of NS 1987A.

This work also benefits from the potential high diagnostic power provided by \lynx. The significant increase in sensitivity with respect to \chandra\ allow us to robustly constrain some cooling models and to infer physical properties about NS 1987A and its envelope. This is true either if the putative NS 1987A is detected by \lynx\ or not. Future \lynx\, observations of SN 1987A will be crucial to shed light on the nature of this compact object.

\newpage
\clearpage
\appendix

\section{Upper limits: other scenarios}

\begin{figure*}[!ht]
    \centering
    \includegraphics[width=\textwidth]{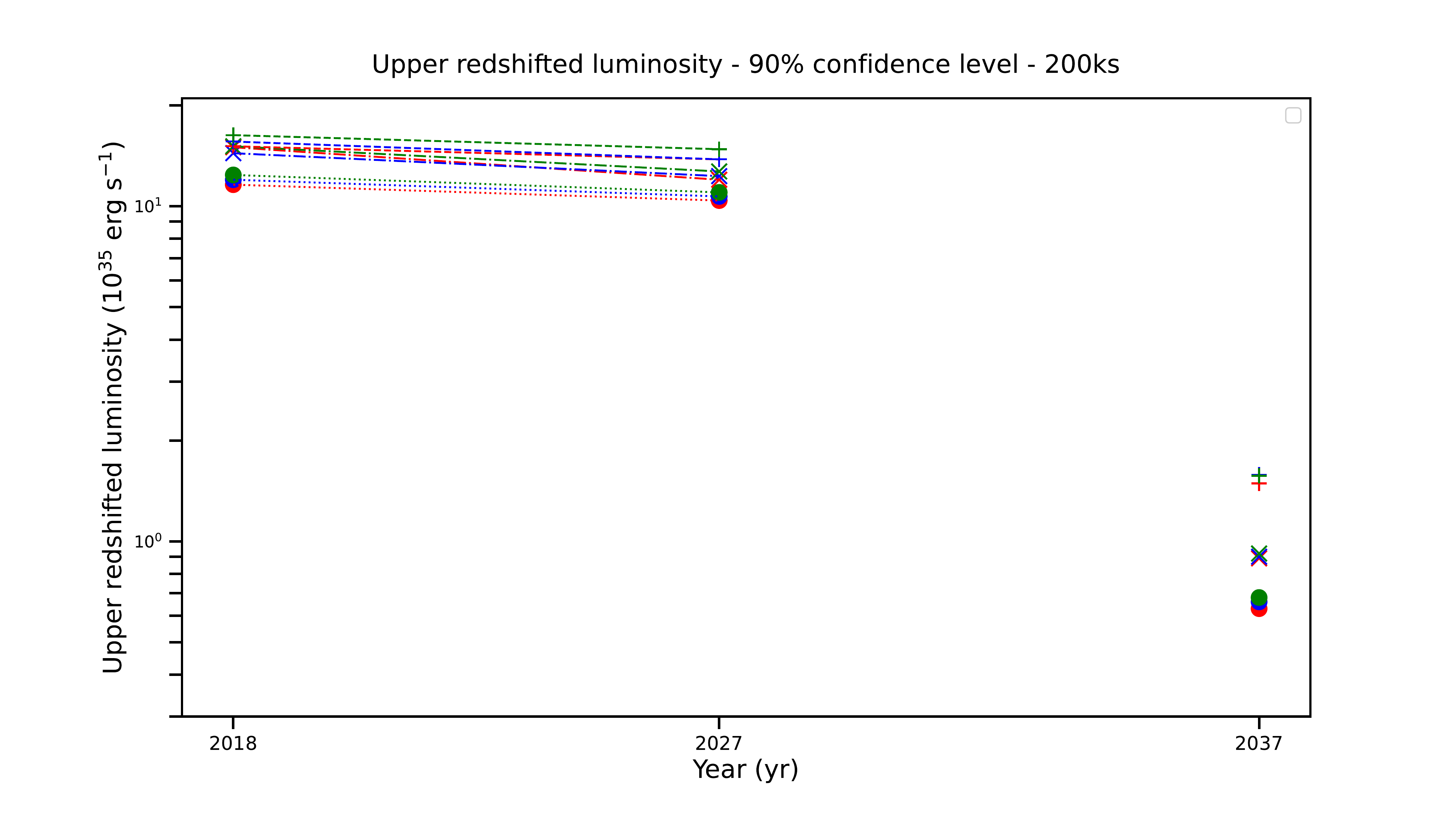}
    \caption{Same as Fig. \ref{fig:upper_limits_90_1Ms} but the synthetic spectra are produced assuming an exposure time of 200 ks.}
    \label{fig:upper_limits_90_200ks}
\end{figure*}

\begin{figure*}[!ht]
    \centering
    \includegraphics[width=\textwidth]{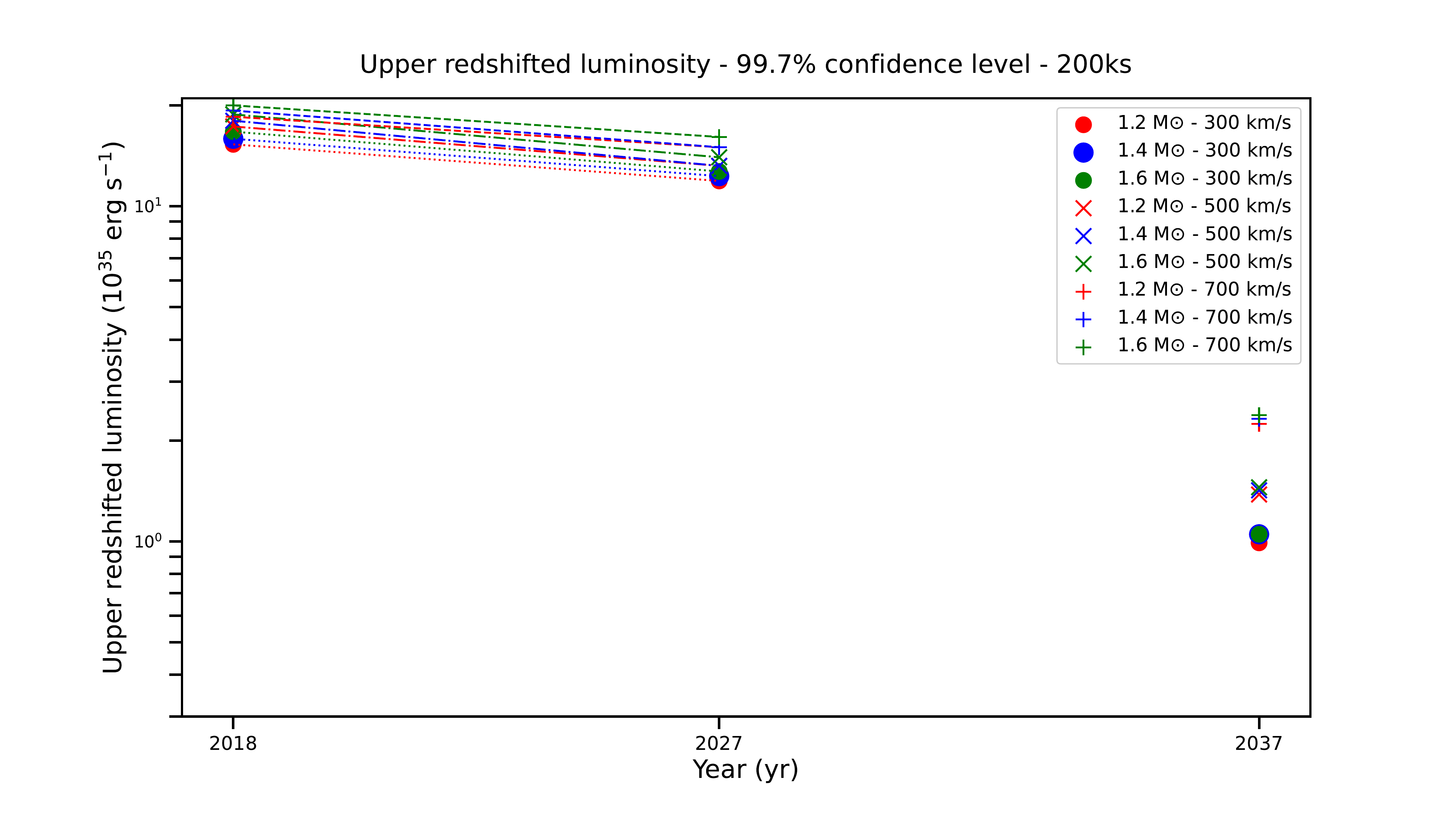}
    \caption{Same as Fig. \ref{fig:upper_limits_90_1Ms} but the upper limits are estimated at the 99.7\% confidence level and the synthetic spectra are produced assuming an exposure time of 200 ks.}
    \label{fig:upper_limits_99_200ks}
\end{figure*}

\begin{figure*}[!ht]
    \centering
    \includegraphics[width=\textwidth]{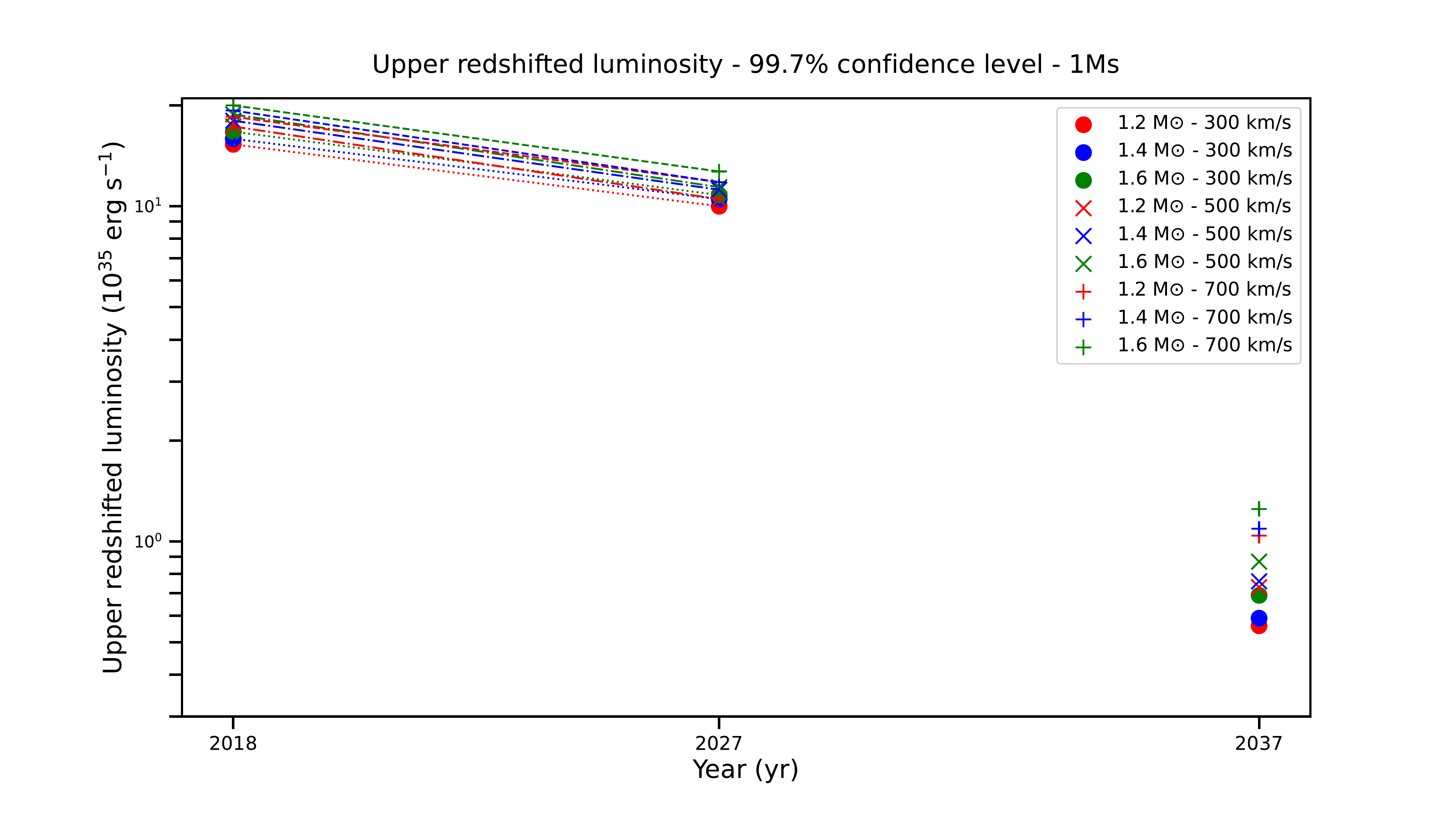}
    \caption{Same as Fig. \ref{fig:upper_limits_90_1Ms} but the upper limits are estimated at the 99.7\% confidence level.}
    \label{fig:upper_limits_99_1Ms}
\end{figure*}

\begin{table}[h]
    \centering
    \caption{Luminosity upper limits for a 1.2M$_{\odot}$ NS}
    \begin{tabular}{c|c|cccc}
    \hline\hline
    No. & Year & $v_{\rm kick}~{\rm (km~s^{-1})}$ & Confidence Level & Exposure time$^b$ (Ms) &Upper limits$^a$ ($10^{35}~{\rm erg~s^{-1}}$) \\
    \hline
    1 &  & 700 & 99.7\% & Table \ref{tab:obs} &18.5 \\
    2 & 2018 & 500 & 99.7\% & Table \ref{tab:obs} &17.3 \\
    3 &  & 300 & 99.7\% & Table \ref{tab:obs} & 15.3 \\
    \hline
    4 &  & 700 & 90\% & Table \ref{tab:obs} & 15.1\\
    5 & 2018 & 500 & 90\% & Table \ref{tab:obs} & 15.0\\
    6 &  & 300 & 90\% & Table \ref{tab:obs} & 11.6 \\
    \hline
    \hline
    7 &  &  700 & 99.7\% & 0.2 & 15.0\\
    8 & 2027 &  500 & 99.7\% & 0.2 & 13.2\\
    9 &   &  300 & 99.7\% & 0.2 & 11.9\\
    \hline
    10 &  &  700 & 90\% & 0.2 & 13.8\\
    11 & 2027 &  500 & 90\% & 0.2 & 12.0\\
    12 &  &  300 & 90\% & 0.2 & 10.4\\
    \hline
    13 &  & 700 & 99.7\% & 1 &  11.8\\
    14 &  2027 & 500 & 99.7\% & 1 & 10.5\\
    15 &  &  300 & 99.7\% & 1 & 10.0\\
    \hline
    16 &  & 700 & 90\% & 1 &  9.88\\
    17 & 2027  &  500 & 90\% & 1 & 9.28\\
    18 &  &  300 & 90\% & 1 & 8.06\\
    \hline
    \hline
    19 &  &  700 & 99.7\% & 0.2 & 2.24\\
    20 & 2037 &  500 & 99.7\% & 0.2 & 1.38\\
    21 &   &  300 & 99.7\% & 0.2 & 0.99\\
    \hline
    22 &  &  700 & 90\% & 0.2 & 1.49\\
    23 & 2037 &  500 & 90\% & 0.2 & 0.89\\
    24 &  &  300 & 90\% & 0.2 & 0.63\\
    \hline
    25 &  & 700 & 99.7\% & 1 & 1.04\\
    26 &  2037 & 500 & 99.7\% & 1 &0.73\\
    27 &  &  300 & 99.7\% & 1 & 0.56\\
    \hline
    28 &  & 700 & 90\% & 1 & 0.65\\
    29 & 2037  &  500 & 90\% & 1 & 0.46\\
    30 &  &  300 & 90\% & 1 & 0.37\\
    \end{tabular}
    
     \textbf{Notes}.     $^a$ Upper limits on the redshifted bolometric luminosities assuming a blackbody with a radius of 11.65 km and a mass of 1.2 M$_{\odot}$. $^b$ Exposure time assumed to synthesize the 2027 \chandra\ and 2037 \lynx\ spectra. 
    \label{tab:upper_limits_1.2}
\end{table}

\begin{table}[h]
    \centering
    \caption{Luminosity upper limits for a 1.4M$_{\odot}$ NS}
    \begin{tabular}{c|c|cccc}
    \hline\hline
    No. & Year & $v_{\rm kick}~{\rm (km~s^{-1})}$ & Confidence Level & Exposure time$^b$ (Ms) &Upper limits$^a$ ($10^{35}~{\rm erg~s^{-1}}$) \\
    \hline
    1 &  & 700 & 99.7\% & Table \ref{tab:obs} & 19.3 \\
    2 & 2018 & 500 & 99.7\% & Table \ref{tab:obs} & 18.0\\
    3 &  & 300 & 99.7\% & Table \ref{tab:obs} & 15.9 \\
    \hline
    4 &  & 700 & 90\% & Table \ref{tab:obs} & 15.6\\
    5 & 2018 & 500 & 90\% & Table \ref{tab:obs} & 14.4\\
    6 &  & 300 & 90\% & Table \ref{tab:obs} & 12.0 \\
    \hline
    \hline
    7 &  &  700 & 99.7\% & 0.2 & 15.0\\
    8 & 2027 &  500 & 99.7\% & 0.2 & 13.2\\
    9 &   &  300 & 99.7\% & 0.2 & 12.3\\
    \hline
    10 &  &  700 & 90\% & 0.2 &13.8\\
    11 & 2027 &  500 & 90\% & 0.2 & 12.3\\
    12 &  &  300 & 90\% & 0.2 & 10.7\\
    \hline
    13 &  & 700 & 99.7\% & 1 &  11.8\\
    14 &  2027 & 500 & 99.7\% & 1 & 11.2\\
    15 &  &  300 & 99.7\% & 1 & 10.5\\
    \hline
    16 &  & 700 & 90\% & 1 &  9.88\\
    17 & 2027  &  500 & 90\% & 1 & 9.55\\
    18 &  &  300 & 90\% & 1 & 8.15\\
    \hline
    \hline
    19 &  &  700 & 99.7\% & 0.2 & 2.32\\
    20 & 2037 &  500 & 99.7\% & 0.2 & 1.42\\
    21 &   &  300 & 99.7\% & 0.2 & 1.05\\
    \hline
    22 &  &  700 & 90\% & 0.2 & 1.58\\
    23 & 2037 &  500 & 90\% & 0.2 & 0.90\\
    24 &  &  300 & 90\% & 0.2 & 0.66\\
    \hline
    25 &  & 700 & 99.7\% & 1 & 1.09 \\
    26 &  2037 & 500 & 99.7\% & 1 & 0.76\\
    27 &  &  300 & 99.7\% & 1 & 0.59\\
    \hline
    28 &  & 700 & 90\% & 1 & 0.69 \\
    29 & 2037  &  500 & 90\% & 1 & 0.49\\
    30 &  &  300 & 90\% & 1 & 0.39\\
    \end{tabular}
    
     \textbf{Notes}.    $^a$ Upper limits on the redshifted bolometric luminosities assuming a blackbody with a radius of 11.57 km and a mass of 1.4 M$_{\odot}$. $^b$ Exposure time assumed to synthesize the 2027 \chandra\ and 2037 \lynx\ spectra. 
    \label{tab:upper_limits_1.4}
\end{table}

\begin{table}[h]
    \centering
    \caption{Luminosity upper limits for a 1.6 M$_{\odot}$ NS}
    \begin{tabular}{c|c|cccc}
    \hline\hline
    No. & Year & $v_{\rm kick}~{\rm (km~s^{-1})}$ & Confidence Level & Exposure time$^b$ (Ms) &Upper limits$^a$ ($10^{35}~{\rm erg~s^{-1}}$) \\
    \hline
    1 &  & 700 & 99.7\% & Table \ref{tab:obs} & 20.0\\
    2 & 2018 & 500 & 99.7\% & Table \ref{tab:obs} & 18.8\\
    3 &  & 300 & 99.7\% & Table \ref{tab:obs} & 16.7\\
    \hline
    4 &  & 700 & 90\% & Table \ref{tab:obs} & 16.3\\
    5 & 2018 & 500 & 90\% & Table \ref{tab:obs} & 15.1\\
    6 &  & 300 & 90\% & Table \ref{tab:obs} & 12.4\\
    \hline
    \hline
    7 &  &  700 & 99.7\% & 0.2 & 16.1\\
    8 & 2027 &  500 & 99.7\% & 0.2 & 14.0\\
    9 &   &  300 & 99.7\% & 0.2 & 12.7\\
    \hline
    10 &  &  700 & 90\% & 0.2 &14.8\\
    11 & 2027 &  500 & 90\% & 0.2 & 12.7\\
    12 &  &  300 & 90\% & 0.2 & 11.0\\
    \hline
    13 &  & 700 & 99.7\% & 1 &  12.7\\
    14 &  2027 & 500 & 99.7\% & 1 & 11.4\\
    15 &  &  300 & 99.7\% & 1 & 10.8\\
    \hline
    16 &  & 700 & 90\% & 1 &  10.7\\
    17 & 2027  &  500 & 90\% & 1 & 9.94\\
    18 &  &  300 & 90\% & 1 & 8.49\\
    \hline
    \hline
    19 &  &  700 & 99.7\% & 0.2 & 2.38\\
    20 & 2037 &  500 & 99.7\% & 0.2 & 1.45\\
    21 &   &  300 & 99.7\% & 0.2 & 1.05\\
    \hline
    22 &  &  700 & 90\% & 0.2 & 1.57\\
    23 & 2037 &  500 & 90\% & 0.2 & 0.92\\
    24 &  &  300 & 90\% & 0.2 & 0.68\\
    \hline
    25 &  & 700 & 99.7\% & 1 & 1.25\\
    26 &  2037 & 500 & 99.7\% & 1 & 0.87\\
    27 &  &  300 & 99.7\% & 1 & 0.69\\
    \hline
    28 &  & 700 & 90\% & 1 & 0.83\\
    29 & 2037  &  500 & 90\% & 1 & 0.60\\
    30 &  &  300 & 90\% & 1 & 0.48\\
    \end{tabular}
    
     \textbf{Notes}. $^a$ Upper limits on the redshifted bolometric luminosities assuming a blackbody with a radius of 11.45 km and a mass of 1.6 M$_{\odot}$. $^b$ Exposure time assumed to synthesize the 2027 \chandra\ and 2037 \lynx\ spectra. 
    \label{tab:upper_limits_1.6}
\end{table}

\section*{Acknowledgements}

We thank Lei Sun for giving some useful advice. We also thank the referee for careful reading and many helpful comments, which significantly improved this paper. A.D., S.N., and M.O. thank supports from RIKEN Interdisciplinary Theoretical and Mathematical Sciences Program (iTHEMS) and Pioneering Program of RIKEN for Evolution of Matter in the Universe (r-EMU). This work is supported by JSPS KAKENHI Grant Number 19H00693, JP21K03545, and 22J10448. M.M., S.O. and B.O. acknowledge financial contribution from the PRIN INAF 2019 grant ``From massive stars to supernovae and supernova remnants: driving mass, energy and cosmic rays in our Galaxy''. EG acknowledges funding from the European Union's Horizon 2020 research and innovation program under grant agreement No. 101004131 (SHARP).

{\it Software:} Gnuplot~\citep{2022gnuplot}, CIAO \citep{ciao}, \href{https://heasarc.gsfc.nasa.gov/docs/software/heasoft/}{HEASOFT}, XSPEC \citep{arn96}, \href{https://sites.google.com/cfa.harvard.edu/saoimageds9/home?authuser=0}{DS9}

{\it Facilities:} \href{https://cxc.harvard.edu/index.html}{Chandra}
\bibliography{ref}{}
\bibliographystyle{aasjournal}



\end{document}